\documentclass[prb,preprint]{revtex4-2}
\usepackage{graphicx}
\usepackage{amsmath}
\usepackage{mathtools}
\usepackage{xcolor}
\usepackage{tikz}

\begin{document}

\title{Oscillatory dependence of tunneling magnetoresistance on barrier thickness
in magnetic tunnel junctions}

\author{B. C. Lee}
\email{chan@inha.ac.kr}
\affiliation{%
 Department of Physics, Inha University\\
 100 Inha-ro, Michuhol-gu, Incheon 22212, Republic of Korea
}%

\date{\today}

\begin{abstract}

The dependence of tunneling conductance and tunneling magnetoresistance (TMR) on barrier thickness in magnetic tunnel junctions is theoretically investigated. The complex band structure of the insulator is taken into account, and an analytical formula for tunneling conductance and TMR is derived. Numerical calculations using a tight-binding model validate the analytical formula. The complex nature of insulator's band structure leads to significant oscillations in tunneling conductance and TMR as functions of barrier thickness. It is demonstrated that these TMR oscillations are not caused by quantum confinement within the barrier, but are instead analogous to classical two-slit optical interference.

\end{abstract}

\vspace{0.5in}

\maketitle

\newpage

\section{Introduction}

In quantum mechanics, tunneling phenomena represent one of the most intriguing 
and non-intuitive aspects of the quantum world.
Tunneling plays a crucial role in numerous physical processes and technologies. 
For instance, it underlies the operation of devices such as the tunnel diode and the scanning tunneling microscope. The tunneling effect has been expanded to 
include spin-dependent transport, as seen in magnetic tunnel junctions (MTJs) \cite{mood}.
A typical MTJ consists of two ferromagnetic (FM)
layers separated by a thin insulating barrier. 
The resistance of MTJ depends on the relative orientation of magnetizations in the FM layers.
When the magnetizations are aligned parallel (P), the resistance is lower 
compared to when they are antiparallel (AP), leading to a measurable change in resistance
known as tunneling magnetoresistance (TMR). 
In the early days, aluminum oxide served as the tunnel barrier.
Replacing amorphous aluminum oxide with crystalline MgO as the tunnel barrier 
resulted in a significant enhancement of TMR \cite{park,yuasa}.
TMR is a fundamental phenomenon in the field of spintronics, and
it has significant implications for various applications, including magnetic sensors
and memory devices like magnetic random-access memory.

Understanding physical properties of TMR is crucial for optimizing 
the performance of MTJs. It allows for the design of spintronic devices with maximum TMR, 
enhancing their efficiency and reliability.
One of the critical factors influencing TMR is the thickness of the insulating 
tunnel barrier.
The barrier thickness directly affects the quantum tunneling probability of electrons
between the two FM layers. As the barrier becomes thicker, the tunneling probability
decreases exponentially, leading to changes in the overall resistance of the MTJ. 
The relationship between barrier thickness and TMR is not straightforward.
While an increase in barrier thickness generally reduces tunneling current, 
it can also enhance spin-dependent tunneling effects under certain conditions, 
potentially increasing TMR. Conversely, if the barrier is too thin, 
direct exchange interactions between the FM layers can suppress TMR 
by reducing the spin-filtering effect.
When the insulating tunnel barrier is crystalline, TMR has been 
experimentally observed to oscillate around a constant value as barrier thickness 
increases \cite{yuasa,mat,maru,Sch21,Sch22,Sch23}.
The resistance of the MTJ not only increases exponentially 
but also oscillates as the barrier thickness increases. 
This adds another intriguing dimension to the tunneling effect, 
as the tunneling probability in a square potential problem is 
typically expected to decrease monotonically as a function of the 
barrier thickness.
Despite theoretical papers on oscillatory TMR \cite{butler,heil,mathon2,zhang},
the origin of these oscillations remains undetermined, 
and the oscillation period has yet to be explained.

Physical properties of TMR are closely linked to the electronic structure of the materials 
involved and the specific quantum states available for tunneling.
In a real insulator, the wavevector component along the conduction direction 
is not purely imaginary but instead a complex number.
The complex band structure is a crucial concept 
in understanding the electronic properties of insulating materials, 
particularly in the context of quantum tunneling and electronic transport. 
This allows for a more detailed analysis of the electronic states within the band gap, 
where no propagating states exist. 
This study aims to investigate how the complex band structure of an insulator
affects the dependence of tunneling conductance and TMR on barrier thickness.
An analytical formula will be derived and validated through numerical calculations.
It will be confirmed that the complex band structure of an insulator indeed leads to significant oscillations in TMR in MTJs.

\section{Analytical Formula for tunneling conductance}

\begin{figure}
\includegraphics[width=12cm, trim=25cm 20cm 3cm 1cm]{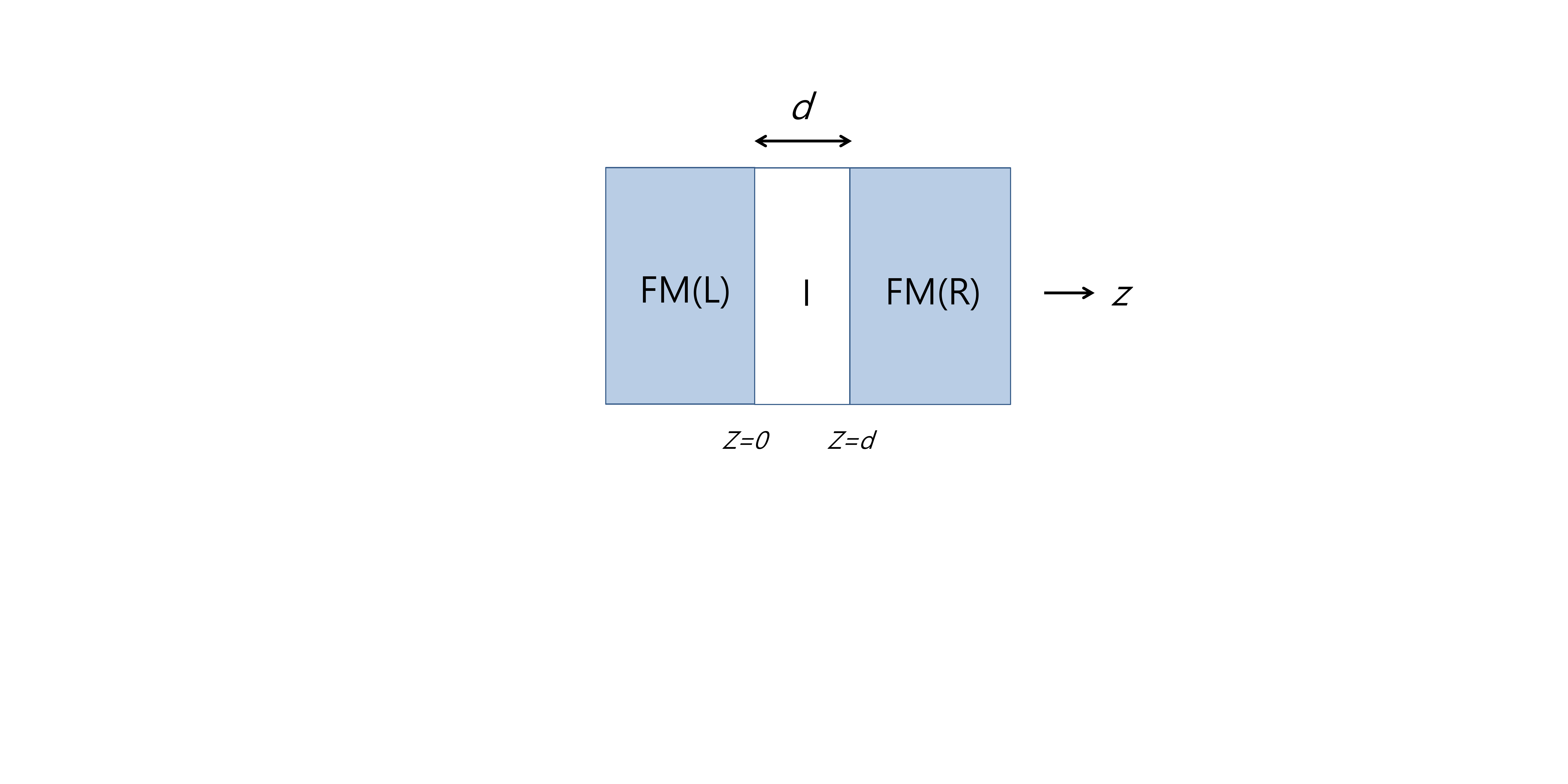}
\caption{ Schematic of a magnetic tunnel junction (MTJ) which consists of the
left ferromagnetic layer [FM(L)], the insulator (I), and 
the right ferromagnetic layer [FM(R)].
$d$ is the thickness of the insulator. 
The $z$-axis is perpendicular to the interface.} \label{MTJ17}
\end{figure}

The structure of a typical MTJ is shown in Fig.~\ref{MTJ17}. The left 
ferromagnetic layer [FM(L)]  and the right ferromagnetic layer [FM(R)] 
are separated by an insulating tunnel barrier (I).
The $z$-axis is defined as the direction of both growth and conduction. 
The thickness of
the tunnel barrier is $d$, and the interface between the FM(L) [FM(R)] 
and the insulator is located at $z=0$ ($z=d$).
It is assumed that the electron spin ($\sigma$) and the wavevector
component parallel to the interface (${\bf{k} }_{\|}$) are
conserved. $2N_{L \sigma}$ ($2N_{R \sigma}$) is the number of bulk states 
in the FM(L)  [FM(R)]
for a given energy $\varepsilon$, ${\bf{k} }_{\|}$, and spin $\sigma$. 
The normalized bulk
solution of the material for the FM(L) is
denoted as $\left| {\bf k}_{\|},k_{z,n \sigma}^{L+(-)}
\right\rangle$ for a given energy $\varepsilon$, 
where $k_z$ is the $z$ component of the wavevector, $n$ is
the band index, $\sigma$ is the spin index, and the $+(-)$ sign is
for the state traveling to the right (left). Similarly, the normalized bulk
solutions of the FM(R) are expressed as $\left|
 {\bf k}_{\|},k_{z,n \sigma}^{R+(-)} \right\rangle$. 
A theoretical approach employed in metal/insulator/metal junctions \cite{lee17} is
modified and further developed for application in MTJs.
The eigenstate of the MTJ can be expressed with a linear combination of 
bulk eigenstates.
When there is only one incoming wave $\left|   {\bf k}_{\|},k_{z,n \sigma}^{L+} \right\rangle$ , 
the corresponding state of the MTJ is written as
\begin{equation}
 \left| \psi_{n \sigma}({\varepsilon,\bf
k}_{\|}) \right\rangle = \left\{
\begin{array}{ll}
\displaystyle \left|  {\bf k}_{\|},k_{z,n \sigma}^{L+}
\right\rangle   +
 \sum_{n=1}^{N_{L \sigma}} \mathcal{R}_{nn',\sigma} \left| 
 {\bf k}_{\|}, k_{z,n' \sigma}^{L-} \right\rangle 
, & z <0,  \\
 \displaystyle  \sum_{n=1}^{N_{R \sigma}}   \mathcal{T}_{nn',\sigma}
                      \left|  {\bf k}_{\|}, k_{z,n' \sigma}^{R+} \right\rangle    , & z > d,
                       \end{array}
                       \right.
\label{eq:MTJ}
\end{equation}
where $\mathcal{T}_{nn',\sigma}$ ($\mathcal{R}_{nn',\sigma}$) is a 
transmission (reflection) amplitude for the incoming wave of 
$\left|  \, {\bf
 k}_{\|},\, k_{z,n \sigma}^{L+} \right\rangle$ to be transmitted (reflected) to
$\left| \, {\bf k}_{\|},\, k_{z,n' \sigma}^{R+}
\right\rangle $ ($\left| \, {\bf k}_{\|},\, k_{z,n'
\sigma}^{L-} \right\rangle $).
$\mathcal{T}_{nn',\sigma }$  and $\mathcal{R}_{nn',\sigma}$ are to be 
determined from the boundary conditions. 
The transmission coefficient for the incoming wave  $\left|  \, {\bf
 k}_{\|},\, k_{z,n \sigma}^{L+} \right\rangle$ to be transmitted to
$\left|  \, {\bf k}_{\|},\, k_{z,n' \sigma}^{R+}
\right\rangle $ is
\begin{equation}
T_{nn',\sigma } = \left| \mathcal{T}_{nn',\sigma} \right|^2
        \frac{I^R_{n',\sigma}}{I^L_{n,\sigma}},
\end{equation}
where $I^L_{n,\sigma}$ ($I^R_{n',\sigma}$) is the incoming (transmitted) current associated 
with  $\left| \, {\bf
 k}_{\|},\, k_{z,n \sigma}^{L+} \right\rangle$
 ($\left|  \, {\bf k}_{\|},\, k_{z,n' \sigma}^{R+}
\right\rangle $).
Conductance ($G$) for low bias
and zero temperature is calculated from the Landauer-B\"{u}ttiker formalism 
as follows \cite{vent}
\begin{equation}
G = \frac{e^2}{h}\sum_{{\bf{k}}_{\|},\sigma} 
     \sum_{n=1}^{N_{L \sigma}} \sum_{n'=1}^{N_{R \sigma}}  
    T_{nn',\sigma} (\varepsilon_F,{\bf k}_{\|})  , 
\label{eq:lan} 
\end{equation}
where $\varepsilon_F$ is the Fermi energy.

\begin{figure}
\begin{tikzpicture}
\draw (2,0) -- (2,3);
\draw[->] (1,1.3) -- (1.9,1.3);
\draw[->] (2.1,1.3) -- (3,1.8);
\draw[->] (2.1,1.3) -- (3,0.8);
\node at (0.5,1.3) {$k^L$};
\node at (3.5,1.8) {$k^{I+}_1$};
\node at (3.5,0.8) {$k^{I+}_2$};
\node at (2.5,2) {$\mathcal{T}^L_1$};
\node at (2.5,0.6) {$\mathcal{T}^L_2$};
\node at (1,3) {FM(L)};
\node at (2.5,3) {I};
\draw (7,0) -- (7,3);
\draw[->] (6,1.8) -- (6.9,1.35);
\draw[->] (6,0.8) -- (6.9,1.25);
\node at (5.5,1.8) {$k^{I+}_1$};
\node at (5.5,0.8) {$k^{I+}_2$};
\draw[->] (7.1,1.3) -- (8,1.3);
\node at (8.5,1.3) {$k^{R}$};
\node at (7.5,1.8) {$\mathcal{T}^R_1$};
\node at (7.5,0.8) {$\mathcal{T}^R_2$};
\node at (8.3,3) {FM(R)};
\node at (6.5,3) {I};
 \end{tikzpicture}
\caption{ Tunneling process through the insulating layer is separated to
three parts: transmission at the FM(L)/I interface,
propagation in the tunnel barrier, and transmission at the I/FM(R) interface.}
\label{MTJsp}
\end{figure}
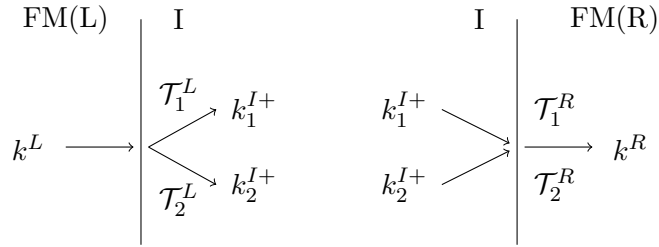

As shown in Fig.~\ref{MTJsp},
the tunneling process in the FM(L)/I/FM(R) tunnel junction can be
separated to three parts: transmission at interface
FM(L)/I, propagation in the barrier, and transmission at interface I/FM(R).
Total transmission coefficient $T_{nn',\sigma}$ can be expressed
with the transmission amplitudes for separated interfaces.
For simplicity, let us consider a scenario where only two exponentially 
decaying states exist within the barrier. 
The more complex multi-band case of the barrier can be reduced 
to this two-state model, as will be explained shortly.
At the FM(L)/I interface,
the incoming wave $\left| {\bf k}_{\|},k^L_{n\sigma} \right\rangle$ is
transmitted to $\left| {\bf k}_{\|},k^{I+}_{1\sigma} \right\rangle$ 
and $ \left| {\bf k}_{\|},k^{I+}_{2\sigma} \right\rangle$,
and the corresponding transmission amplitudes are denoted as
$\mathcal{T}^L_{n,1\sigma}$ and $\mathcal{T}^L_{n,2\sigma} $,
respectively. In the barrier,
$\left| {\bf k}_{\|},k^{I+}_{1\sigma} \right\rangle$ and
$\left| {\bf k}_{\|},k^{I+}_{2\sigma} \right\rangle$
are two available states which decay to the right.
The incoming wave $\left| {\bf k}_{\|},k^L_{n\sigma} \right\rangle$ is 
spin-dependent, and states $\left| {\bf k}_{\|},k^{I+}_{1\sigma} \right\rangle$ and
$\left| {\bf k}_{\|},k^{I+}_{2\sigma} \right\rangle$
in the barrier are also spin-dependent, in general, 
although the barrier is nonmagnetic.
Inside the tunnel barrier, the wavefunction can be expressed as
\begin{equation}
    \left| \psi^I_{\sigma}(\varepsilon, {\bf k}_{\|}  \right  \rangle
    =  e^{ik^{I+}_{1\sigma} z} \mathcal{T}^L_{n,1\sigma} 
      \left| {\bf k}_{\|},k^{I+}_{1\sigma} \right\rangle
      +  e^{ik^{I+}_{2\sigma} z} \mathcal{T}^L_{n,2\sigma} 
      \left| {\bf k}_{\|},k^{I+}_{2\sigma} \right\rangle .
\label{eq:psiI}
\end{equation}
The wavefunction is decaying inside the barrier, and the imaginary parts of 
$k^{I+}_{1\sigma}$ and $k^{I+}_{2\sigma}$ should be positive.
Even when there are more electronic bands in the barrier, 
the wavefunctions with larger Im($k^{I+}_{1\sigma}$) decay faster
and can be ignored as barrier thickness increases. 
Thus, the two-state model
in Eq.~(\ref{eq:psiI}) can be applied to most of cases.
The two transmitted wavefunctions in the barrier are orthogonal 
and do not interfere each other.
At the right I/FM(R) interface, both wavefunctions $\left| {\bf k}_{\|},k^{I+}_{1\sigma} \right\rangle$ 
and $ \left| {\bf k}_{\|},k^{I+}_{2\sigma} \right\rangle$ are transmitted to wavefunction $\left| {\bf k}_{\|},k^R_{n'\sigma} \right\rangle $. 
The corresponding transmission amplitudes are $\mathcal{T}^R_{1\sigma,n'} $
and $\mathcal{T}^R_{2\sigma,n'} $, respectively.
In short, the wavefunction in the FM(L) is separated
into two wavefunctions inside the barrier, and they interfere
upon arriving at the FM(R) after the tunneling. 
The total transmission amplitude
$\mathcal{T}_{nn',\sigma}$ is expressed as 
\begin{equation}
    \mathcal{T}_{nn',\sigma} 
  =   \mathcal{T}^R_{1\sigma,n'} e^{ik^{I+}_{1\sigma} d} \mathcal{T}^L_{n,1\sigma} 
           + \mathcal{T}^R_{2\sigma,n'} e^{ik^{I+}_{2\sigma} d} \mathcal{T}^L_{n,2\sigma} .
\end{equation}
There are multiple reflection inside the tunnel barrier, 
but the contribution from the multiple reflection to the tunneling
current is negligible because the wavefunction decays exponentially
in the insulator. The total transmission
coefficient $T_{nn',\sigma}$ is
${T}_{nn',\sigma} =\left| \mathcal{T}_{\sigma} \right|^2
        I^R_{n',\sigma} / I^L_{n,\sigma}$, where $I^L_{n,\sigma}$ ($I^R_{n',\sigma}$)
is the current in the  FM(L) [FM(R)]. The total 
transmission coefficient $T_{nn',\sigma}$ is expressed with the
transmission amplitudes of the separated interfaces such as,
\begin{eqnarray}
   {T}_{nn',\sigma}
  &=& \left[ e^{-2{\rm Im}(k^{I+}_{1\sigma}) d} \left|  \mathcal{T}^L_{n,1\sigma} \mathcal{T}^R_{1\sigma,n'} \right|^2
           +e^{-2{\rm Im}(k^{I+}_{2\sigma}) d}\left|\mathcal{T}^L_{n,2\sigma} \mathcal{T}^R_{2\sigma,n'} \right|^2 \right]  
           \frac{I^R_{n',\sigma}}  {I^L_{n,\sigma}} \nonumber \\
  & & + 2 {\rm Re} \left[ e^{i(k^{I+}_{1\sigma} - k^{I+*}_{2\sigma})d}  
     \mathcal{T}^L_{n,1\sigma} \mathcal{T}^R_{1\sigma,n'} 
    \mathcal{T}^{L*}_{n,2\sigma} \mathcal{T}^{R*}_{2\sigma,n'}  \right] 
     \frac{I^R_{n',\sigma}}  {I^L_{n,\sigma}} .
     \label{eq:trans}
\end{eqnarray}
The total transmission can be divided to two parts: 
the directly squared terms and the cross term.
The directly squared terms are simply decaying exponentially. 
The cross term also decays exponentially due to 
the imaginary part of $k^{I+}_{1 \sigma} - k^{I+}_{2 \sigma}$,
and may be oscillatory at same time 
when $k^{I+}_{1 \sigma} - k^{I+}_{2 \sigma}$ has a non-zero real part.
Within the barrier, the wavefunction decays exponentially, and quantum confinement does not occur. 
Consequently, the oscillation of the total transmission coefficient, which results in TMR oscillation, 
is unrelated to the quantum confinement effect. 
The primary mechanism behind TMR oscillation as a function of barrier thickness is 
fundamentally different from other oscillatory TMR effects in MTJs \cite{zeng,yang,niu,choi}.
The two wavefunctions within the barrier represent different eigenstates and do not interfere with 
each other inside the barrier. 
Interference only occurs at the interface, once the two tunneling waves reach the the FM(R). 
This situation is similar to optical two-slit interference. 
If only a single transmitted state exists in the barrier, there will be no oscillatory behavior, 
even if the z-component of the wavevector has a non-zero real part.
$k^{I+}_{1 \sigma}$ and $k^{I+}_{2 \sigma}$ can be obtained from the energy
dispersion relation $\varepsilon_F=\varepsilon({\bf{k}}_{\|},k^I_{z \sigma})$
of the barrier. 
Assume that the insulator has the reflection symmetry, 
which means $\varepsilon_F=\varepsilon({\bf{k}}_{\|},k^I_{z \sigma})
=\varepsilon({\bf{k}}_{\|},-k^I_{z \sigma})$.
When the complex conjugate is taken for the energy dispersion relation,
$\varepsilon_F=\varepsilon({\bf{k}}_{\|},k^{I*}_{z \sigma})$ is obtained.
In the two-band case with the reflection symmetry, if $k^{I+}_{1 \sigma}$
is expressed as $k^{I+}_{1 \sigma} = k_{r \sigma}^I + i k_{i \sigma}^I$ 
($k_{r \sigma}^I$ and $k_{i \sigma}^I$ are real and positive),
$k^{I+}_{2 \sigma}$ is automatically 
$k^{I+}_{2 \sigma} = -k_{r \sigma}^I + i k_{i \sigma}^I$.
${\rm Im}(k^{I+}_{1\sigma})={\rm Im}(k^{I+}_{2\sigma})=k_{i \sigma}^I $  
and $k^{I+}_{1 \sigma} -k^{I+*}_{2 \sigma}= 2k_{r \sigma}^I + 2i k_{i \sigma}^I$ are obtained,
and both the direct squared terms and the interference term decay exponentially 
as a function of the barrier thickness, following a $e^{-2 k_{i \sigma}^I d}$ pattern.
Simultaneously, the interference term oscillates as $e^{2i k_{r \sigma}^I d}$, 
with an oscillation period of $\pi/k_{r \sigma}^I$. 
From this point forward, I will assume reflection symmetry for the insulator.
The conductance is described by the Landauer-B\"{u}ttiker formalism in 
Eq.~(\ref{eq:lan}). 
In general, multiple traveling states within the FM layer contribute to conduction. 
I will focus on the one band that most closely matches with the band in the barrier 
and dominates tunneling conduction, omitting the band indices $n$ and $n'$ for simplicity. 
If there is significant contribution from other bands in the FM layer, 
they can be easily incorporated using a similar calculation.
The conductance is obtained by summing 
transmission coefficient $T_{\sigma}$ over ${\bf{k}}_{\|}$,
and main contribution comes from the
vicinity of a ${\bf{k}}_{\|}$-point where $ k_{i \sigma}^I$ is at its minimum 
and, consequently, $T_{\sigma}$ is at a maximum. 
Let us assume that $ k_{i \sigma}^I$ is the smallest
at ${\bf{k}}_{\|}={\bf{k}}_{\| 0}=(k_{x0},k_{y0})$,
and denote the minimum value as $ k_{i0 \sigma}^I$ 
[$ k_{i0 \sigma}^I = k_{i \sigma}^I ({\bf{k}}_{\|}={\bf{k}}_{\| 0}) $].
Also assume that $ k_{r \sigma}^I$ has an extremal value
at ${\bf{k}}_{\|}={\bf{k}}_{\| 0}$ with 
$ k_{r0 \sigma}^I = k_{r \sigma}^I ({\bf{k}}_{\|}={\bf{k}}_{\| 0}) $. 
$ k_{i \sigma}^I$  is approximated as 
\begin{eqnarray}
k^{I}_{i \sigma} ({\bf{k}}_{\|}) \cong k^{I}_{i0 \sigma} + \frac{1}{2}\frac{(k_x-k_{x0})^2}{\kappa_{ix \sigma}}
+ \frac{1}{2}\frac{(k_y-k_{y0})^2}{\kappa_{iy \sigma}}
\end{eqnarray}
with
$\frac{1}{\kappa_{ix  \sigma}} = \left. \frac{\partial^2 k_{i  \sigma}}{\partial k_x^2}  \right|_{{\bf{k}}_{\|}={\bf{k}}_{\| 0}}$ and
$\frac{1}{\kappa_{iy  \sigma}} = \left. \frac{\partial^2 k_{i  \sigma}}{\partial k_y^2}  \right|_{{\bf{k}}_{\|}={\bf{k}}_{\| 0}}$.
Similarly, $ k_{r \sigma}^I$ is expressed as
\begin{eqnarray}
k^{I}_{r \sigma} ({\bf{k}}_{\|}) \cong k^{I}_{r0 \sigma} + \frac{1}{2}\frac{(k_x-k_{x0})^2}{\kappa_{rx \sigma}}
+ \frac{1}{2}\frac{(k_y-k_{y0})^2}{\kappa_{ry \sigma}}
\end{eqnarray}
with
$\frac{1}{\kappa_{rx  \sigma}} = \left. \frac{\partial^2 k_{r  \sigma}}{\partial k_x^2}  \right|_{{\bf{k}}_{\|}={\bf{k}}_{\| 0}}$ and
$\frac{1}{\kappa_{ry  \sigma}} = \left. \frac{\partial^2 k_{r  \sigma}}{\partial k_y^2}  \right|_{{\bf{k}}_{\|}={\bf{k}}_{\| 0}}$.
Integration over ${\bf{k}}_{\|}$ is carried out analytically,
and the conductance per unit area is 
\begin{equation}
G   =
   \sum_{\sigma}\left[ A_{\sigma \sigma'} \frac{e^{-2k^{I}_{i0 \sigma} d}  }{d} 
        +\left| B_{\sigma \sigma'} \right| \frac{e^{-2k^{I}_{i0 \sigma} d}  }{d} 
          \cos (2 k_{r0\sigma}^I d + \phi_{\sigma \sigma'}) \right]  ,
\end{equation}
where constants $A_{\sigma}$, $ B_{\sigma}$, and $\phi_{\sigma}$ are
\begin{eqnarray}
     A_{\sigma \sigma'} &=&  \frac{e^2 \sqrt{\kappa_{ix \sigma} \kappa_{iy \sigma}}}{4 \pi h} \left(
     \left| \mathcal{T}^L_{1 \sigma}  \mathcal{T}^R_{1 \sigma'} \right|^2 + \left|  \mathcal{T}^L_{2 \sigma}\mathcal{T}^R_{2 \sigma'}  \right|^2 \right) 
   \frac{I^R_{\sigma'}}  {I^L_{\sigma}}, \nonumber \\
   B_{\sigma \sigma'} &=& 
   \frac{e^2 \sqrt{\kappa_{ix \sigma} \kappa_{iy \sigma}}}{2 \pi h} 
    \sqrt{ \frac{\kappa_{rx \sigma}\kappa_{ry \sigma}}
                {(\kappa_{rx \sigma} -i\kappa_{ix \sigma}) 
                (\kappa_{ry \sigma} -i\kappa_{iy \sigma}) } }
       \mathcal{T}^L_{1 \sigma} \mathcal{T}^R_{1 \sigma'} 
    \mathcal{T}^{L*}_{2 \sigma}  \mathcal{T}^{R*}_{2 \sigma'} 
     \frac{I^R_{\sigma'}}  {I^L_{\sigma}},  \nonumber
\end{eqnarray}
and $ \phi_{\sigma \sigma'} = \arg ({B_{\sigma \sigma'})}$.
All the parameters such as $ \mathcal{T}^R_{1 \sigma}$ and 
$ \mathcal{T}^L_{1 \sigma}$ are calculated at ${\bf{k}}_{\|}={\bf{k}}_{\| 0}$ and the Fermi level $\varepsilon = \varepsilon_F$.
$\sigma$ ($\sigma'$) is the electron spin in the 
FM(L) [FM(R)] layer.
The absolute direction of spin $\sigma'$ is same as that of $\sigma$.
The electron spin with a given direction may be either majority or
minority spin in the FM layer depending on the direction of the magnetization.
The majority (minority) spin of the electron is denoted as 
$\uparrow$ ($\downarrow$). 
When the magnetization of the two FM layers is parallel,
the conductance per unit area is
\begin{eqnarray}
    G_P  & = & 
    A_{\uparrow \uparrow} \frac{e^{-2k^{I}_{i0 \uparrow} d}  }{d} 
          +  A_{\downarrow \downarrow} \frac{e^{-2k^{I}_{i0 \downarrow} d}  }{d}  \nonumber \\
          & &
        +\left| B_{\uparrow \uparrow} \right| \frac{e^{-2k^{I}_{i0 \uparrow} d}  }{d} 
          \cos (2 k_{r0\uparrow}^I d + \phi_{\uparrow \uparrow})  
        +\left| B_{\downarrow \downarrow} \right| \frac{e^{-2k^{I}_{i0 \downarrow} d}  }{d} 
          \cos (2 k_{r0\downarrow}^I d + \phi_{\downarrow \downarrow}).  
\end{eqnarray}  
When there are other kinds of $k^{I}_{i0 \uparrow}$ and
$k^{I}_{i0 \downarrow}$ corresponding to different ${\bf{k}}_{\| 0}$
in the multiple-band case, each 
channel related to $k^{I}_{i0 \uparrow}$ or
$k^{I}_{i0 \downarrow}$ contributes to the conductance, and
it needs to be summed over all possible $k^{I}_{i0 \uparrow}$ and
$k^{I}_{i0 \downarrow}$.
Similarly, the conductance per unit area for the
AP magnetization is
\begin{eqnarray}
    G_{AP}  & = & 
    A_{\uparrow \downarrow} \frac{e^{-2k^{I}_{i0 \uparrow} d}  }{d} 
          +  A_{\downarrow \uparrow} \frac{e^{-2k^{I}_{i0 \downarrow} d}  }{d}  \nonumber \\
          & &
        +\left| B_{\uparrow \downarrow} \right| \frac{e^{-2k^{I}_{i0 \uparrow} d}  }{d} 
          \cos (2 k_{r0\uparrow}^I d + \phi_{\uparrow \downarrow})  
        +\left| B_{\downarrow \uparrow} \right| \frac{e^{-2k^{I}_{i0 \downarrow} d}  }{d} 
          \cos (2 k_{r0\downarrow}^I d + \phi_{\downarrow \uparrow}).  
\end{eqnarray}  
Summation over all possible $k^{I}_{i0 \uparrow}$ and
$k^{I}_{i0 \downarrow}$ is implied in this case, too.

The conductance decays exponentially as the barrier thickness $d$
increases. When there are many different kinds of 
$k^{I}_{i0 \uparrow}$ and $k^{I}_{i0 \downarrow}$,
conductance is dominated by contribution associated with the
smallest $k^{I}_{i0\uparrow}$ or $k^{I}_{i0 \downarrow}$,
provided that the corresponding $ \mathcal{T}^R$ and 
$ \mathcal{T}^L$ are not negligible.
The barrier is nonmagnetic and the smallest value of 
$k^{I}_{i0 \uparrow}$ or $k^{I}_{i0 \downarrow}$ is spin-independent.
The smallest is denoted as $k^{I}_{i0}$ and
corresponding real part of $k^I_z$ is expressed as $k^{I}_{r0}$.
The conductance per unit area for P magnetization is
\begin{eqnarray}
    G_P  & = & \frac{e^{-2k^{I}_{i0 } d}  }{d} 
    \left[
   A_{\uparrow \uparrow} +A_{\downarrow \downarrow} 
      +\left| B_{\uparrow \uparrow} + B_{\downarrow \downarrow} \right| 
          \cos (2 k_{r0}^I d + \phi_{P}) \right],  
\label{eq:Gp}
\end{eqnarray}  
where $\phi_{P}$ is $\phi_{P} = \arg(B_{\uparrow \uparrow} + B_{\downarrow \downarrow})$.
The conductance per unit area for AP magnetization is
\begin{eqnarray}
    G_{AP}  & = & 
  \frac{e^{-2k^{I}_{i0 } d}  }{d} 
    \left[
   A_{\uparrow \downarrow} +A_{\downarrow \uparrow} 
      +\left| B_{\uparrow \downarrow} + B_{\downarrow \uparrow} \right| 
          \cos (2 k_{r0}^I d + \phi_{AP}) \right] , 
\label{eq:Gap}
\end{eqnarray}  
where $\phi_{AP}$ is $\phi_{AP} = \arg(B_{\uparrow \downarrow} + B_{\downarrow \uparrow})$.
A general form of the conductance per unit area as a function of 
the thickness $d$ is
\begin{equation}
    G =  \frac{e^{-2k^{I}_{i0 } d}}{d} [A + B \cos(2 k_{r0}^I d + \phi)],
\label{eq:conductance}
\end{equation}
where $A$, $B$, and $\phi$ can be considered as fitting parameters
in experiments.
The resistance multiplied by the sample area, $RA$, for the MTJ
is expressed as
\begin{equation}
    RA =  \frac {d e^{2k^{I}_{i0 } d}}{A + B \cos(2 k_{r0}^I d + \phi)}.
\end{equation}
The resistance $R$ increases as $d e^{2k^{I}_{i0 } d}$ 
and also oscillates with period $\pi/k_{r0}$ as the barrier thickness $d$ increases. 
$k^{I}_{i0 }$ and $k^I_{r0}$ are related to the 
complex Fermi surface of the barrier. 
$k^{I}_{i0 }$ is the shortest vector from the origin to
the imaginary part of the complex Fermi surface along the growth direction.
As will be explained later, $2 k_{r0}$ is the extremal 
spanning wavevector of the real part of 
the complex Fermi surface in the conduction direction.
The TMR is $(G_P - G_{AP})/G_{AP}$. It neither decays nor grows, 
but oscillates with increasing the barrier thickness $d$.

\section{Tight-binding band calculation of tunneling
conductance}

\begin{figure}
\includegraphics[width=12cm, trim=0 6cm 3cm 0, clip]{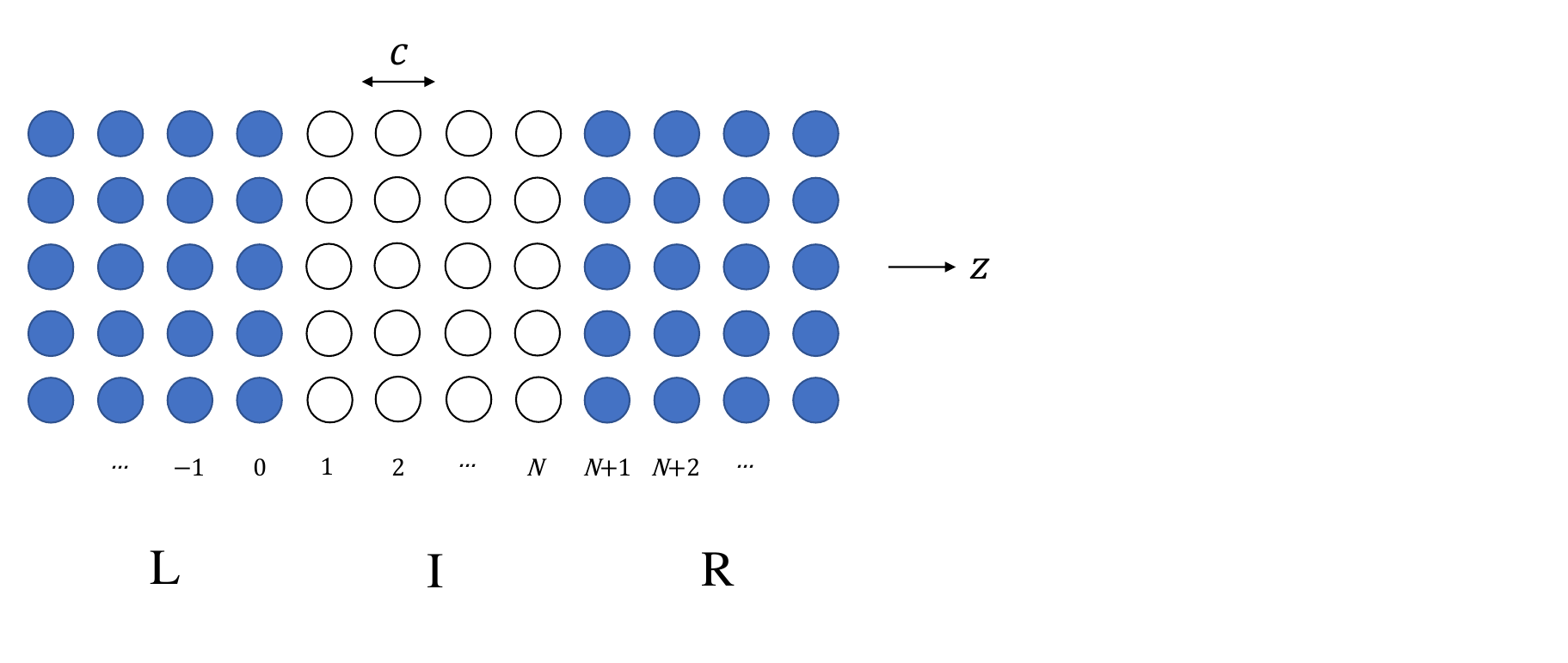}
\caption{ Magnetic tunnel junction (MTJ) which consists of a
left (L) magnetic layer, the insulator (I), and right (R) magnetic layer.
A simple cubic lattice structure is taken for each layer.
$c$ is the lattice constant and also the thickness of the atomic plane. 
Each atomic plane perpendicular to the $z$-axis is labelled by a number at the bottom.} \label{MTJ}
\end{figure}

A numerical calculation is carried out to examine the analytical formula
derived in Sec.~II. 
It is assumed that each material has a simple cubic
structure as shown in Fig.~\ref{MTJ}.
The growth direction is taken as the $z$-axis, and
$c$ is the atomic constant. 
There are $N$ atomic planes in the tunnel barrier, and
the barrier thickness $d$ is $d=Nc$.
A tight-binding (TB) scheme employed in metal/insulator/metal junctions \cite{lee21}
is modified for the band calculation in this paper, and
only the nearest-neighbor interaction is considered.
The atomic orbital centered at $ {\bf R}_{i} $ is denoted as
$ \left| {\bf R}_{i} ; \alpha , Z
\right\rangle $, where ${\bf R}_{i}$ is the position vector of an atom
and $Z$ is the material index ($Z=$L, I, or R).


%
The FM material is described by a single-band TB scheme.
The TB parameters for the FM(L) [FM(R)] are
$$\left\langle {\bf R}_{i} ; L(R) \right| H_{L(R)} \left| {\bf
R}_{i} ; L(R) \right\rangle = E_{L(R)} $$  
and
$$\left\langle {\bf R}_{i} ; L(R) \right| H_{L(R)} \left| {\bf
R}_{j} ; L(R) \right\rangle = t_{L(R)} $$
if ${\bf R}_{i}$ is the nearest neighbor of
${\bf R}_{j}$.
$H_{L(R)}$ is the bulk Hamiltonian for the  the FM(L) [FM(R)], and
the orbital index is ignored because there is only one band in the FM layer.
For convenience, it is assumed that the FM(L) and FM(R) 
are made of the same material ($E_M \equiv E_L = E_R$ and $t_M \equiv t_L = t_R$).
The dispersion relation for the FM material is
\begin{equation}
\varepsilon (\mathbf{k}) = E_M + 2t_M \left(\cos{k_x c} + \cos{k_y c} +\cos{k_z c}\right),
\label{eq:d1}
\end{equation}
where $\mathbf{k}=(k_x , k_y, k_z)$ is a wave vector.
Two-spin channel model is adopted for the transport, and 
TB parameters $E_M$ and $t_M$ are spin-dependent
although the spin index is not shown.

For the insulator, a two-band TB model is
employed, and  the two atomic orbitals are denoted as $a$ and $b$. 
It is assumed that the two atomic orbitals are orthogonal.
The TB parameters for the insulator are
$$\left\langle {\bf R}_{i} ; \alpha , I \right| H_I \left| {\bf
R}_{i} ; \alpha' , I \right\rangle = E_{\alpha} \delta_{\alpha
\alpha'} \enspace (\alpha, \alpha' = a, b)$$  
and
$$\left\langle {\bf R}_{i} ; \alpha \right| H_I \left| {\bf
R}_{j} ; \alpha' \right\rangle = t_{\alpha \alpha'} $$
if ${\bf R}_{i}$ is the nearest neighbor of
${\bf R}_{j}$.
$H_I$ is the bulk Hamiltonian for the insulator, and
TB parameters $t_{aa}$ and $t_{bb}$ are
abbreviated as $t_a$ and $t_b$, respectively. 
$t_{ab}$ is taken as a real number ($t_{ab}=t_{ba}$).
The bulk eigenstate of the insulator can be expressed as 
$\left| \psi_{\bf k}^I
\right\rangle = C_a \left| {\bf k};a, I \right\rangle + C_b \left| {\bf
k};b, I \right\rangle$,
where $\left| {\bf k};\alpha, I \right\rangle$ is the Bloch state for
$\alpha$-orbital in the insulator. 
The bulk Hamiltonian of the insulator can be expressed with a matrix form such as
\begin{equation}
H_I
=\left( \begin{array}{cc} E_a + 2t_a \left(\cos{k_x c} + \cos{k_y c} +\cos{k_z c}\right)
& 2t_{ab} \left(\cos{k_x c} + \cos{k_y c} +\cos{k_z c}\right) \\
2t_{ab} \left(\cos{k_x c} + \cos{k_y c} +\cos{k_z c} \right)  &  
E_b + 2t_b \left(\cos{k_x c} + \cos{k_y c} +\cos{k_z c}\right)
\end{array}  \right) .
\label{eq:rt2}
\end{equation}

The wave function of the MTJ can be expressed with
the planar orbital state $\left| {\bf k}_{\|}, l;\alpha , Z \right\rangle$,
which is defined as
\begin{equation}
\left| {\bf k}_{\|},l;\alpha,Z \right\rangle = 
\frac{1}{\sqrt{ N_{\|}}} 
\sum_{i=1}^{ N_{\|}} e^{i{\bf k}_{\|}\cdot {\bf R}_{i \|}} 
\left| {\bf R}_{i \|}, l; \alpha, Z \right\rangle ,
\end{equation}
where $N_{\|}$ is the number of atoms in a given plane, ${\bf k}_{\|}$ is the component of the wave
vector parallel to the plane, ${\bf R}_{i \|}$ 
is the lattice position projected on the given plane, and $l$ is the layer position.
 It is assumed that the planar
atomic-orbital bases are orthogonal, i.e. $\left\langle {\bf
k}_{\|}, l;\alpha \right. \left| {\bf k}_{\|}, l';\alpha'
\right\rangle = \delta_{ll'} \delta_{\alpha \alpha'}$.
The wave function of the MTJ is expressed with bulk states
\begin{eqnarray}
\left| \psi (\varepsilon , {\bf k}_{\|}) \right\rangle &=& \sum_{l= -\infty}^{0} e^{ik^L (l-1/2)c}
\left| {\bf k}_{\|},l;L \right\rangle 
\nonumber \\
& &  + \mathcal{R} \sum_{l= -\infty}^{0}
e^{-ik^L (l-1/2)c} \left| {\bf k}_{\|},l;L \right\rangle  
\nonumber \\
& &  + \sum_{j}\mathcal{D}_{j} \sum_{l= 1}^{N}  e^{ik^I_j (l-1/2)c} \sum_{\alpha}
C_{j,\alpha}\left| {\bf k}_{\|},l;\alpha,I \right\rangle 
  \\
& & + \mathcal{T}
\sum_{l= N+1}^{\infty} e^{ik^R (l-N-1/2)c} \left| {\bf k}_{\|},l;R
\right\rangle  \nonumber ,
\end{eqnarray}
where $k^L$, $k_j^I$, and $k^R$ are
the $z$-components of the wavevector in the FM(L), the insulator,
and the FM(R), respectively, for given energy $\varepsilon$ and ${\bf k}_{\|}$.
Coefficients $\mathcal{R}$, $\mathcal{D}$, and $\mathcal{T}$ are determined
from the boundary conditions, and
the transmission coefficient is $ T = \left| \mathcal{T} \right|^2  \frac{I_R }{I_L } $,
where $I_R$ and $I_L$ are current in the FM(R) and FM(L), respectively. 

 The boundary conditions are obtained by
 projecting the planar states very at the interface
 to the Schr\"{o}dinger equation $ \left( H-\varepsilon \right) \left|
\psi (\varepsilon , {\bf k}_{\|}) \right\rangle  =0$, where $H$
is the Hamiltonian of the MTJ. Only the nearest-neighbor interaction is included, and the planar states with $l=0,1,N,N+1$ are involved. 
The boundary conditions at the left interface are
$\left\langle {\bf k}_{\|},0;L \right| \left( H-\varepsilon \right) \left|
\psi \right\rangle =0$ and
$\left\langle {\bf k}_{\|},1;\alpha,I \right| \left( H-\varepsilon \right)
\left| \psi \right\rangle =0$ ($\alpha = a,b$).
Similarly, the boundary conditions at the right interface are
$\left\langle {\bf k}_{\|},N;\alpha,I \right| \left( H-\varepsilon \right)
\left| \psi \right\rangle =0$ ($\alpha = a,b$) and
$\left\langle {\bf k}_{\|},N+1;R \right| \left( H-\varepsilon \right) \left|
\psi \right\rangle =0$.
Six unknown coefficients $\mathcal{R}$, $\mathcal{D}_1$, 
$\mathcal{D}_2$, $\mathcal{D}_3$, $\mathcal{D}_4$, and $\mathcal{T}$
are to be determined from the boundary conditions.
The first boundary condition, 
$\left\langle {\bf k}_{\|},0;L \right| \left( H-\varepsilon \right) \left|
\psi \right\rangle =0$, leads to \cite{boundary}
\begin{eqnarray}
- \mathcal{R} e^{-\frac{1}{2}ik^L c} \overline{H}_{L}^{(+1)}
+\sum_{j} \mathcal{D}_j e^{ik_{j}(1/2)c}\sum_{\alpha}C_{j,\alpha} 
\overline{H}_{L,I\alpha}^{(+1)} =e^{\frac{1}{2}ik^L c} \overline{H}_{L}^{(+1)},
\end{eqnarray}
where the matrix element $\overline{H}_{Z\alpha,
Z'\alpha'}^{(\sigma)} ({\bf k}_{\|})$ is defined as
\begin{equation}
\overline{H}_{Z\alpha, Z' \alpha'}^{(\sigma)} ({\bf k}_{\|})=
\left\langle {\bf k}_{\|}, l;\alpha,Z \right| \left( H-\varepsilon \right)
\left| {\bf k}_{\|}, l+ \sigma;\alpha' , Z'\right\rangle.
\end{equation}
Eventually, the first boundary conditions is expressed as 
%
\begin{equation}
\begin{aligned}
- e^{-\frac{1}{2}ik^L c} t_L \mathcal{R} 
+  e^{\frac{1}{2}ik^I_{z}c} 
(t_{L,a}C_a + t_{L,b}C_b) \mathcal{D}_1
+ e^{-\frac{1}{2}ik^I_{z}c} 
(t_{L,a}C_a + t_{L,b}C_b) \mathcal{D}_2     \\
+  e^{\frac{1}{2}i{k_{z}^I}^*c} 
(t_{L,a}C_a{}^* + t_{L,b}{C_b}^*) \mathcal{D}_3
+  e^{-\frac{1}{2}i{k_{z}^I}^*c} 
(t_{L,a}C_a{}^* + t_{L,b}{C_b}^*) \mathcal{D}_4
 =e^{\frac{1}{2}ik^L c} t_L  ,
 \end{aligned}
\end{equation}
where $t_{L,a}$ and $t_{L,b}$ are the TB hopping parameters
between two different atoms at the interface.
The other boundary conditions are treated in a similar way.
This method can be extended to the multiple-band case for real materials.


\section{Numerical Results}

\begin{figure}
\includegraphics[width=6cm]{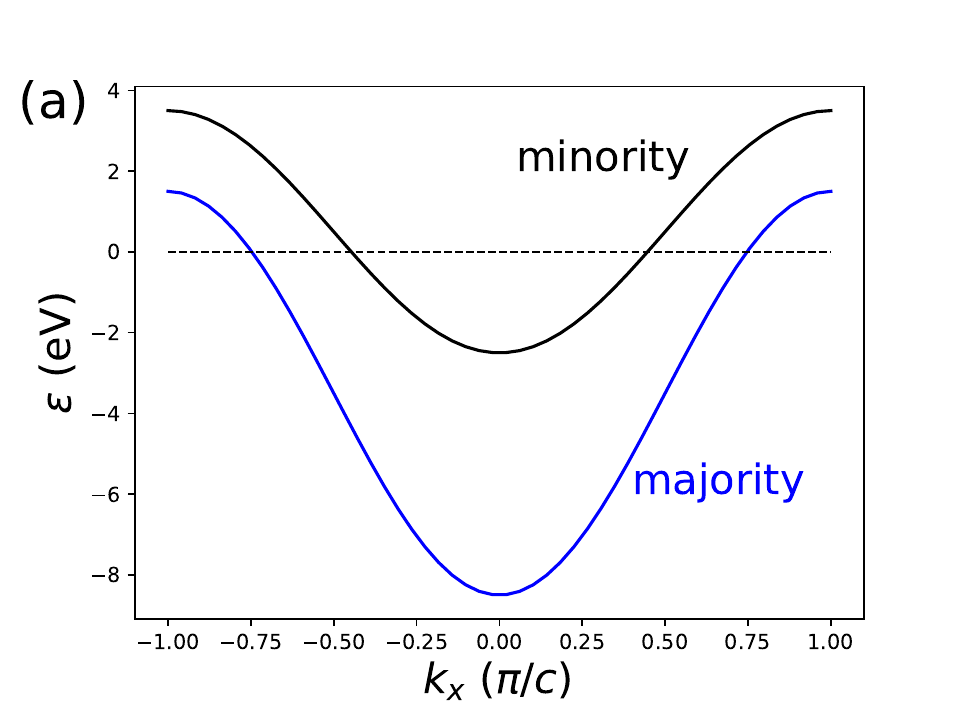}
\includegraphics[width=6cm]{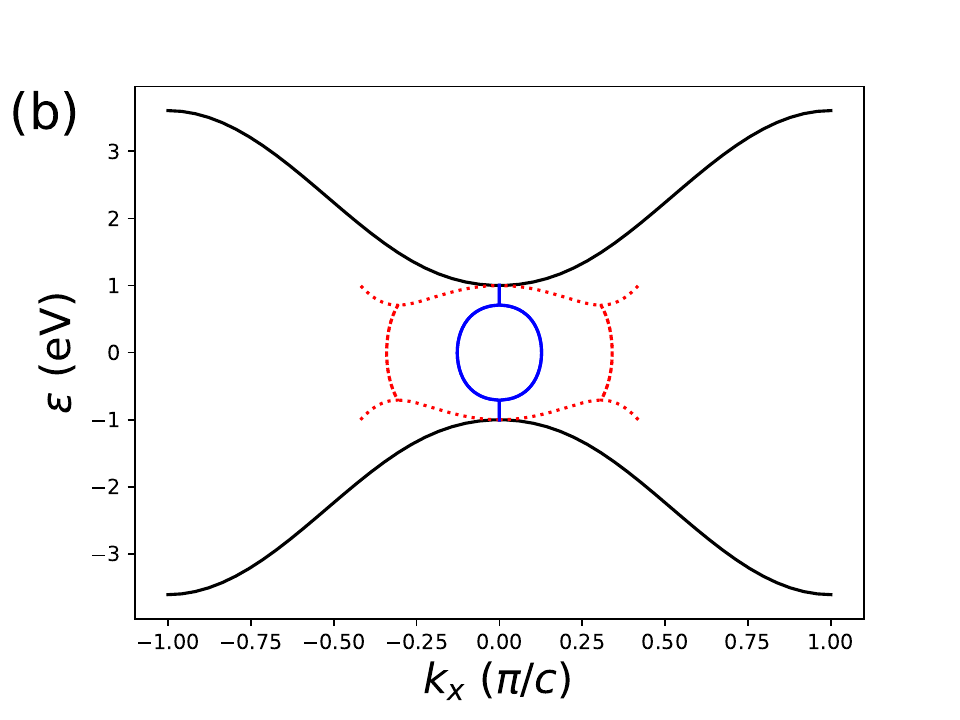}
\includegraphics[width=6cm]{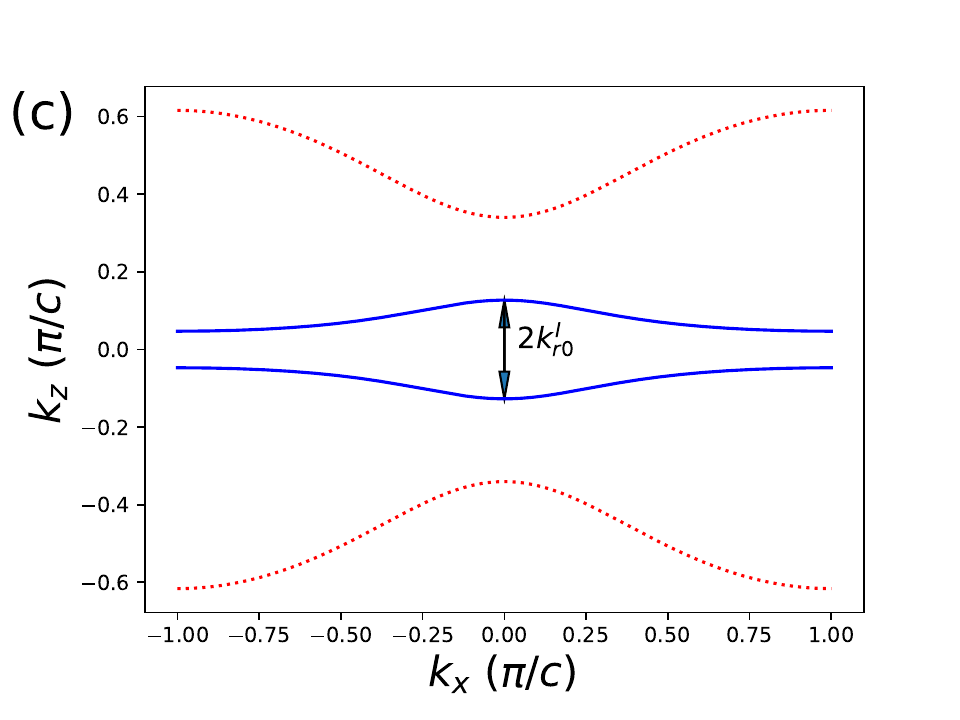}
\includegraphics[width=6cm]{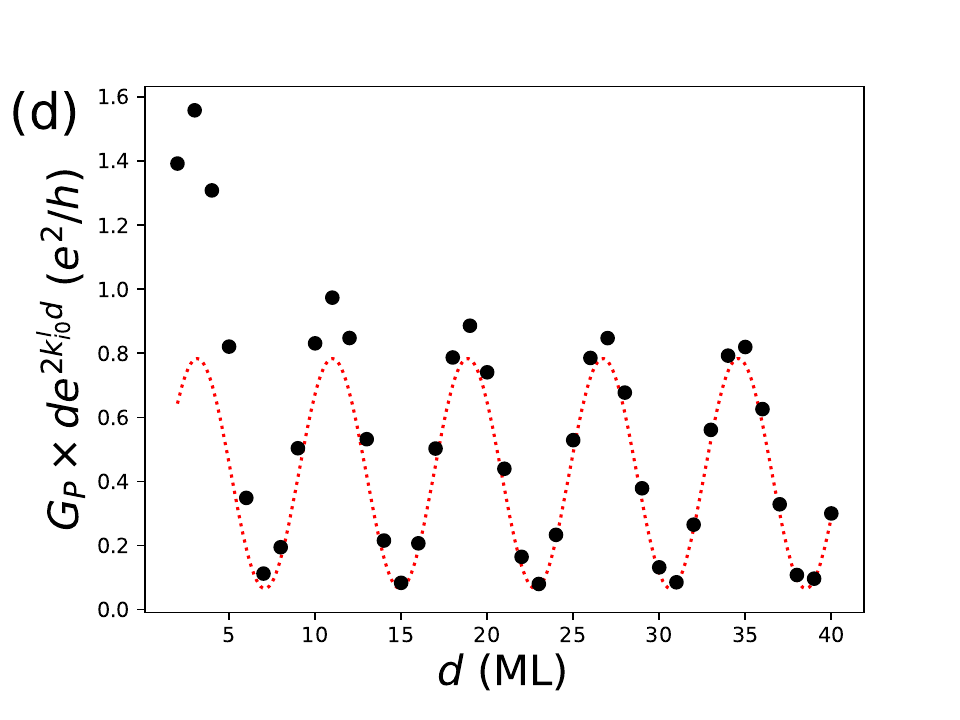}
\includegraphics[width=6cm]{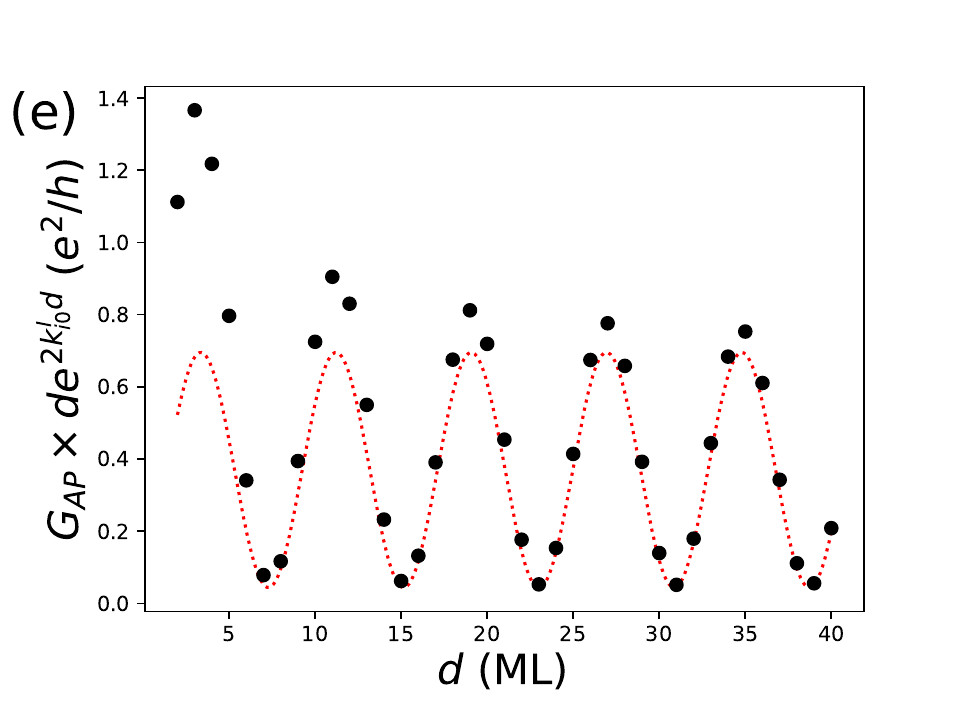}
\includegraphics[width=6cm]{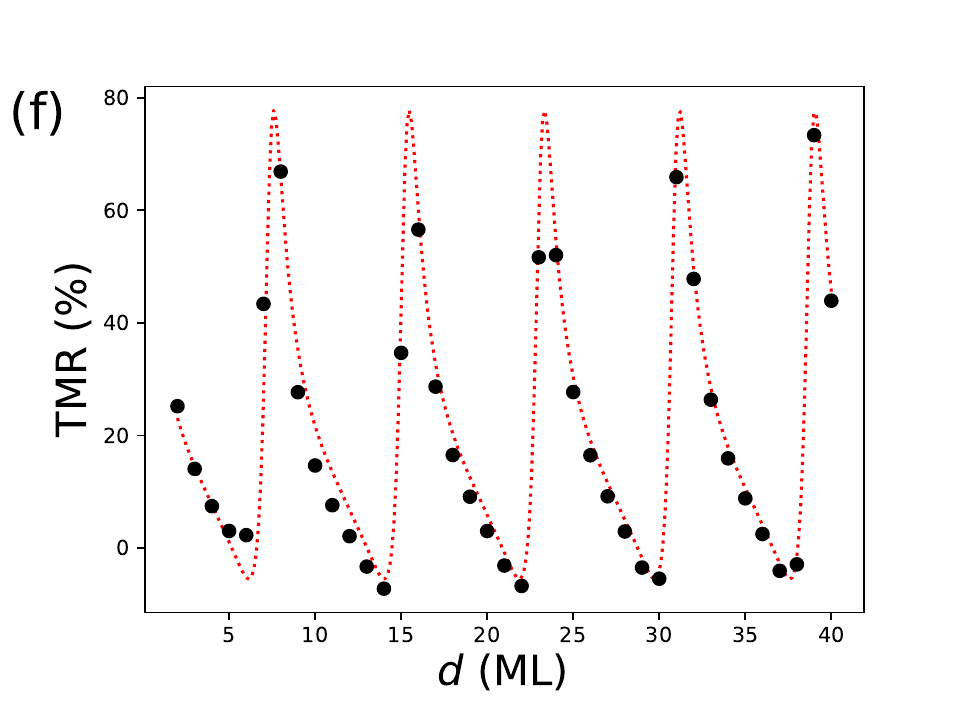}
\caption{ (a) Energy dispersion relation of the FM material
with $k_y = k_z =0$. The TB
parameters are $E_{M\uparrow}=6.5$ eV, $t_{M\uparrow}=-2.5$ eV,
$E_{M\downarrow}=6.5$ eV, and $t_{M\downarrow}=-1.5$ eV in Eq.~(\ref{eq:d1}).
The dotted line is the Fermi level ($\varepsilon_F =0$).
$c$ is the lattice constant.
(b) Complex-band structure of the insulator with $k_y = k_z =0$.
The solid (dotted) line is the real (imaginary) 
part of $k_x$ for given energy. 
The TB parameters are $E_a=-5$ eV, $t_a=0.7$ eV,
$E_b=5$ eV, $t_b=-0.7$ eV, and $t_{ab}=0.1$ eV.
(c) Cross section of the complex Fermi surface along the
$k_x$ axis with $k_y=0$. The solid (dotted) line is the real 
(imaginary) part of the complex Fermi surface.
$2k^I_{r0}$ is the extremal spannig vector of the real part
of the complex Fermi surface.
(d) Tunneling conductance of the MTJ for the parallel
magnetization as a function of the barrier thickness $d$. 
The solid circles are the exact full-band calculation,
and the dotted line is the result of the analytical formula.
(e) Tunneling conductance for the antiparallel
magnetization. The solid circles are the exact full-band calculation,
and the dotted line is the result of the analytical formula.
(f) Dependence of TMR on barrier thickness $d$. 
The solid circles are results of the full-band calculation, and
the dotted line is the result of the analytical formula.
 } \label{TMR1}
\end{figure}

The tunneling conductance and TMR in the MTJ are calculated numerically 
by using the TB band structure described in Sec.~III and 
compared with the results of the analytical formula derived in Sec.~II.
In Fig.~\ref{TMR1}(a), energy ($\varepsilon$) is plotted as a function of
$k_x$ with $k_y=k_z =0$ for a FM material. The upper (lower) band is
for the minority (majority) spin. The Fermi level is set as 
$\varepsilon_F =0$. The Fermi wavevector along [100] orientation is
$k_{F,x}=0.75 \pi/c$ and $0.44 \pi/c$ 
for the majority and minority spin, respectively. 
The energy dispersion relation of the insulator is plotted 
in Fig.~\ref{TMR1}(b). A bandgap exists between $-1$ eV and 1 eV,
with no traveling states at the Fermi level $\varepsilon_F =0$.
In the bandgap, $k_x$ is a complex number, and the real (imaginary)
part of $k_x$ is displayed in the solid (dotted) curve.
The cross section of the complex Fermi surface for the insulator is shown
in Fig.~\ref{TMR1}(c). The solid (dotted) line is the real (imaginary)
part of $k_z$ for the complex Fermi surface. 
$k^I_i$ is the imaginary part of $k_z$, and it has a 
minimum value at $k_x = 0$. Most of 
contribution to the conductance is from tunneling associated with the
extremal point ${\bf k}_{\|}=0$ where $k^I_i$ is the smallest. 
$k^{I}_{i0 }$ denotes the minimum value of
$k^I_i$, and $k^{I}_{i0 }$ is $k^{I}_{i0 }=k^I_i({\bf k}_{\|}=0)$
in this case.
In Eq.~(\ref{eq:conductance}), the tunneling conductance decays exponentially
such as $e^{-2k^{I}_{i0}d}/d$. 
$k^I_{r0}$ is the maximum value of the real part of  $k_z$ for
the complex Fermi surface.
$2 k^I_{r0}$ is the extremal (maximum in this case)
spanning wavevector of the real part of the complex Fermi surface
along the growth direction, as shown in Fig.~\ref{TMR1}(c).  
$2 k^I_{r0}$ is responsible for the oscillation 
part of the tunneling conductance. 
Figures~\ref{TMR1}(d) and (e) show the tunneling conductance 
$G_P$ and $G_{AP}$ as functions of
the barrier thickness $d$ for P and AP magnetization,
respectively. 
The solid circles are the exact numerical results from the 
TB calculation, and the dotted line is the result
of the analytical formula.  
Note that $G \times d e^{2k^{I}_{i0 }d}$ is a
simply oscillatory function,  which is $A + B \cos(2 k_{r0}^I d + \phi)$
in Eq.~(\ref{eq:conductance}).
The analytical formula closely matches the exact numerical calculation, 
particularly as the barrier thickness $d$ increases.
This suggests that the function $e^{-2k^{I}_{i0 }i d}/d$ describes 
the conductance decay more accurately than a simple exponential decay.
The oscillation period is $\pi/ k^I_{r0}$ as predicted by the the analytical formula.
In Fig.~\ref{TMR1}(f), the TMR is plotted as a function of the tunnel
barrier thickness $d$. 
The solid circles are the results of the full-band calculation,
and the dotted line is from the analytical formula.
The analytical formula is in good agreement with the exact numerical calculation.
The TMR oscillates with the same period $\pi/ k^I_{r0}$ as conductance, 
and its amplitude remains constant without decaying or growing. 
Unlike the conductance oscillation, the TMR oscillation is not sinusoidal.
In the analytical formula for TMR, cosine functions appear in both the numerator and denominator, 
leading to a non-sinusoidal TMR. 
The peak-to-valley difference of TMR oscillation is about  80 \%, 
indicating that 
the complex band of the barrier can cause significant TMR oscillations 
as a function of barrier thickness. 

\begin{figure}
\includegraphics[width=6cm]{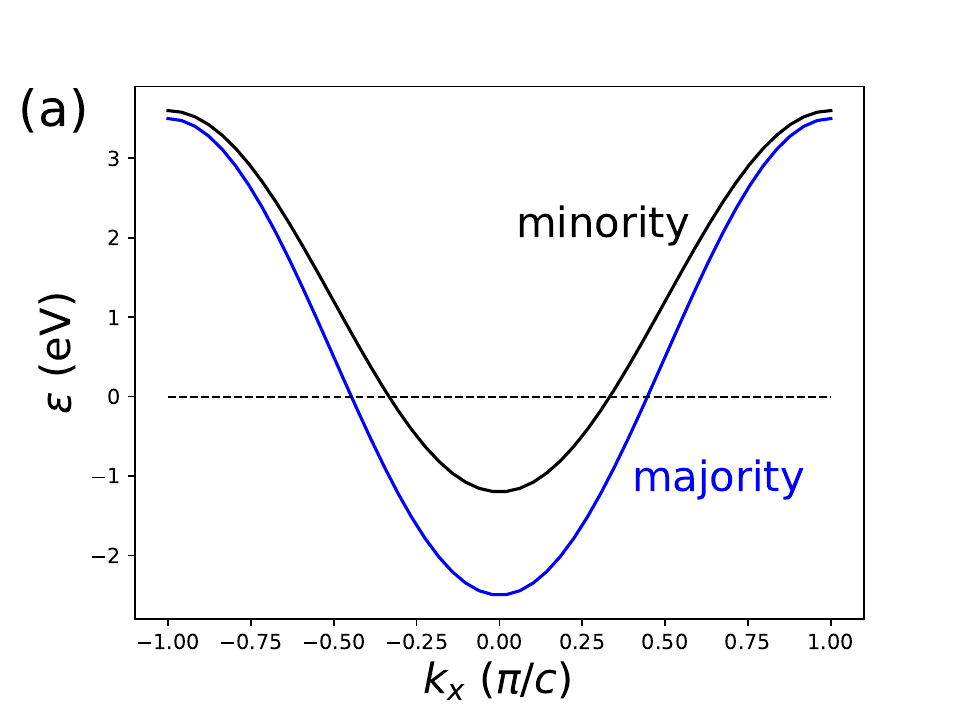}
\includegraphics[width=6cm]{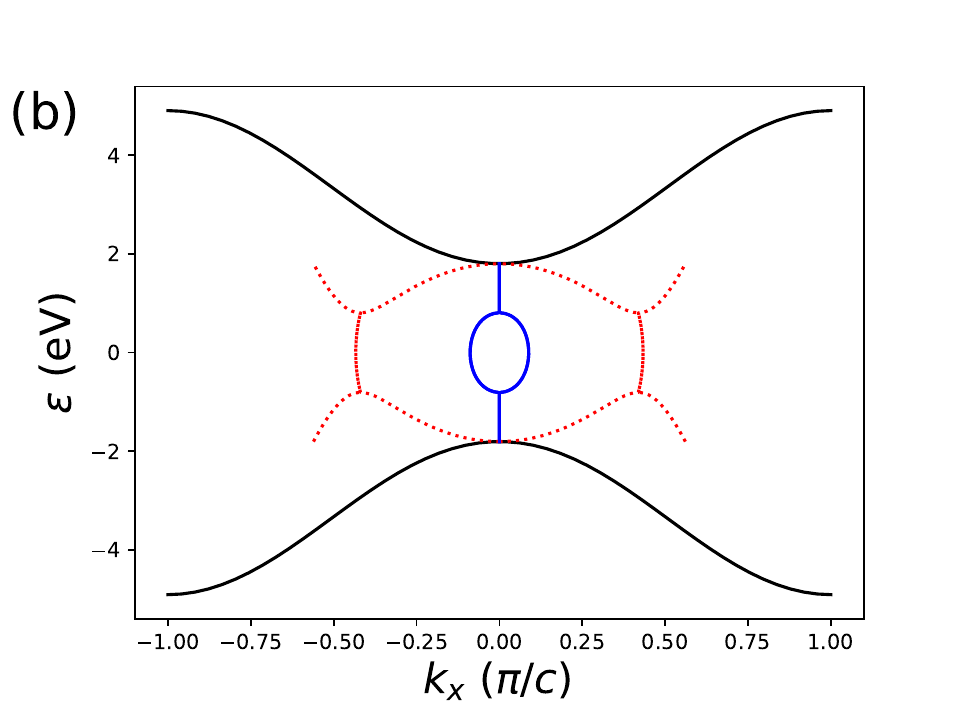}
\includegraphics[width=6cm]{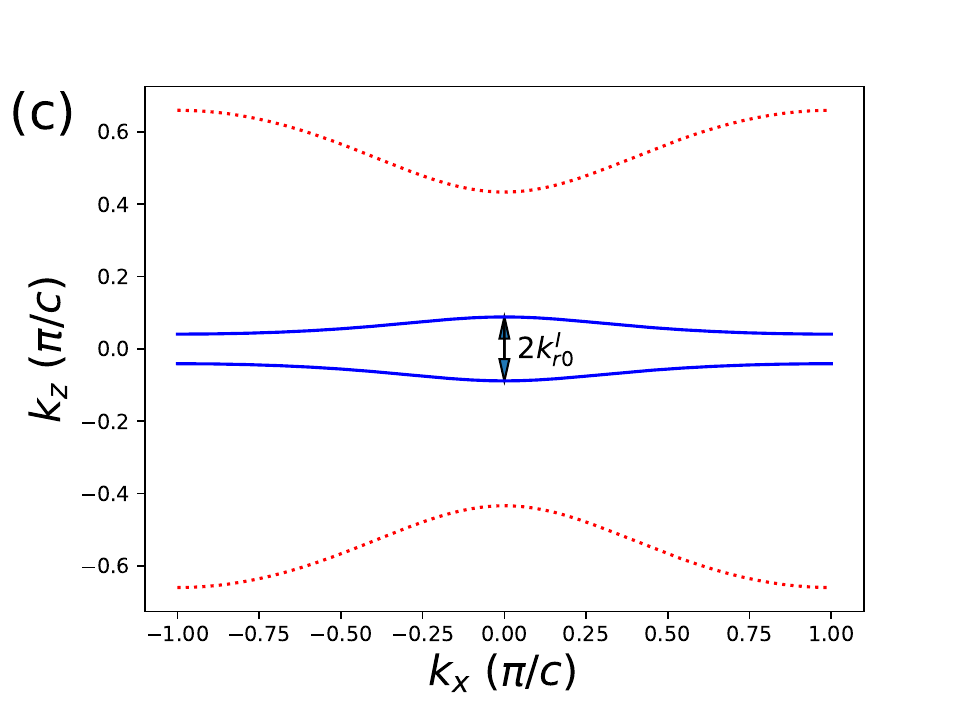}
\includegraphics[width=6cm]{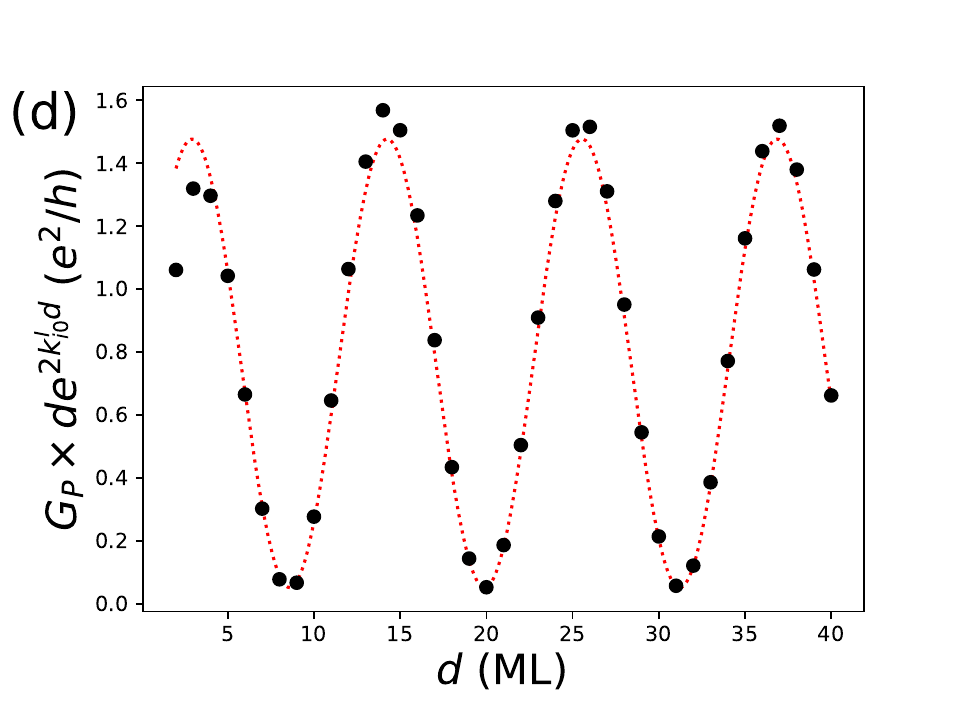}
\includegraphics[width=6cm]{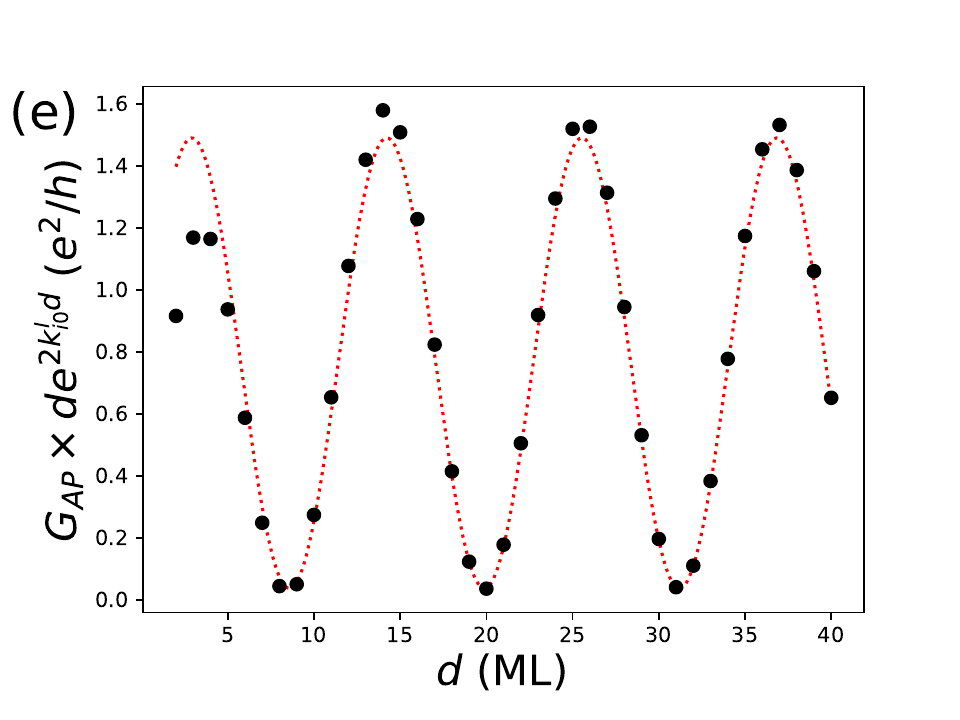}
\includegraphics[width=6cm]{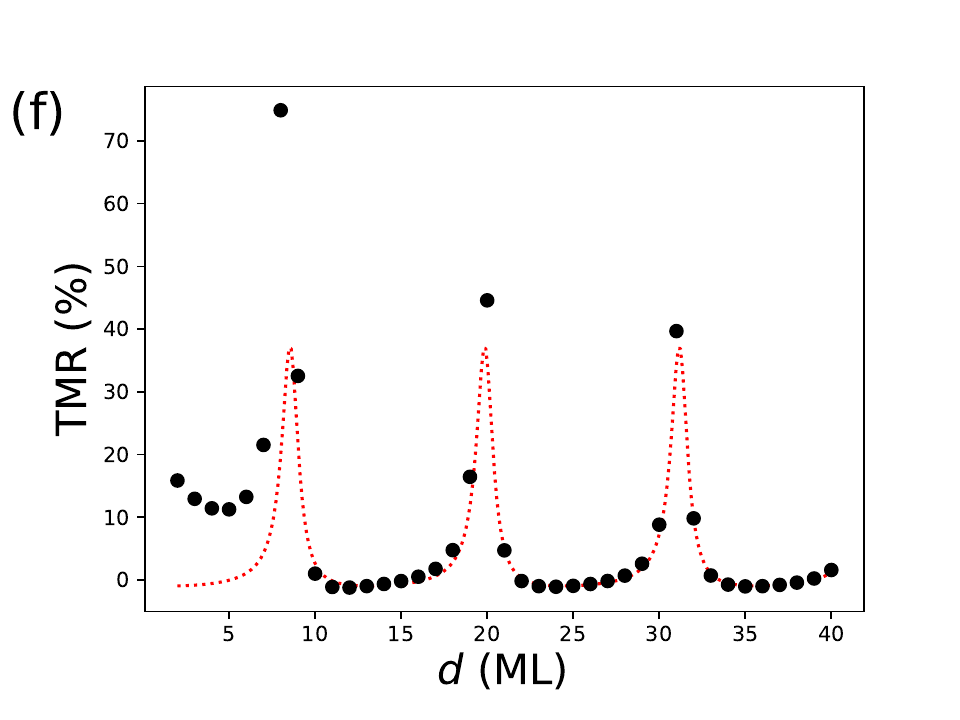}
\caption{ (a) Energy dispersion relation of the FM material
with $k_y = k_z =0$. The TB
parameters are $E_{M\uparrow}=6.5$ eV, $t_{M\uparrow}=-1.5$ eV,
$E_{M\downarrow}=6$ eV, and $t_{M\downarrow}=-1.2$ eV.
(b) Complex band structure of the insulator with $k_y = k_z =0$.
The solid (dotted) line is the real (imaginary) 
part of $k_x$ for given energy. 
The TB parameters are $E_a=-6.5$ eV, $t_a=0.8$ eV,
$E_b=6.5$ eV, $t_b=-0.8$ eV, and $t_{ab}=0.1$ eV.
(c) Cross section of the complex Fermi surface along the
$k_x$ axis with $k_y=0$. The solid (dotted) line is the real 
(imaginary) part of the complex Fermi surface.
(d) Tunneling conductance of the MTJ for the parallel
magnetization as a function of the barrier thickness $d$. 
The solid circles are the exact full-band calculation,
and the dotted line is the result of the analytical formula.
(e) Tunneling conductance of the MTJ for the antiparallel
magnetization. The solid circles are the exact full-band calculation,
and the dotted line is the result of the analytical formula.
(f) TMR as a function of the barrier thickness $d$. 
The black circles are results of the full-band calculation, and
the dotted line is the result of the analytical formula.
 } \label{TMR2}
\end{figure}

For comparison, numerical calculations have been performed 
for similar band structures with different TB parameters, as shown in Fig.~\ref{TMR2}.
The single-band structure of the FM material is shown in Fig.~\ref{TMR2}(a).
The Fermi wavevector along [100] orientation is
$k_{F,x}=0.45\pi/c$ and $0.33\pi/c$ 
for the majority and minority spin, respectively. 
The Fermi surface is considerably smaller than that in Fig.~\ref{TMR1}(a).
The energy dispersion relation of the insulator in Fig.~\ref{TMR2}(b)
is similar to the previous case.
The cross section of the complex Fermi surface is shown in Fig.~\ref{TMR2}(c).
The imaginary part of  $k_{z}$ is the smallest at ${\bf k}_{\|}=0$ 
[$k^{I}_{i0 }=k^I_i({\bf k}_{\|}=0)$].
The extremal spanning wavevector of the real part of
the complex Fermi surface is $2k^{I}_{r0 }$.
The tunneling conductance is plotted as a function of the barrier thickness
for P magnetization in Fig.~\ref{TMR2}(d) and
for AP magnetization in Fig.~\ref{TMR2}(e).
The oscillation period is $\pi/k^I_{r0}$ which is longer than that in Fig.~\ref{TMR1}
since $2k^{I}_{r0 }$ is smaller in this case.
The analytical formula agrees well with the exact calculation.
The agreement is noticeable even at small $d$.
The tunneling probability is the highest at the extremal point ${\bf k}_{\|}=0$.
When deriving the analytical formula, tunneling electrons near the extremal point 
${\bf k}_{\|}=0$ are treated accurately. However, for small $d$,
the tunneling probability as a function of ${\bf k}_{\|}$
may remain significant away from the extremal point, reducing the accuracy of the analytical formula. 
In Fig.~\ref{TMR2}, the Fermi surface of the FM layer is smaller, 
so the tunneling electrons are concentrated closer to the extremal point 
${\bf k}_{\|}=0$, leading to better agreement with the analytical formula.
The TMR is plotted as a function of $d$ in Fig.~\ref{TMR2}(f).
The oscillation period is $\pi/k^I_{r0}$ as expected.
The TMR oscillation has peculiar shape which is totally different
from that in Fig.~\ref{TMR1}(f). 
$A$ is a positive constant in Eq.~(\ref{eq:conductance}).
There are multiple bands in real FM materials, and
conductance contributed by each band needs to be added.
When each contribution is added, positive constant $A$  
increases faster than the oscillation
amplitude $B$ because the phase factor $\phi$ is different
for each contribution.
As a result, the ratio $B/A$ in  Eq.~(\ref{eq:conductance})
will decrease, making the TMR oscillation more rounded and increasingly sinusoidal in shape.

\begin{figure}
\includegraphics[width=6cm]{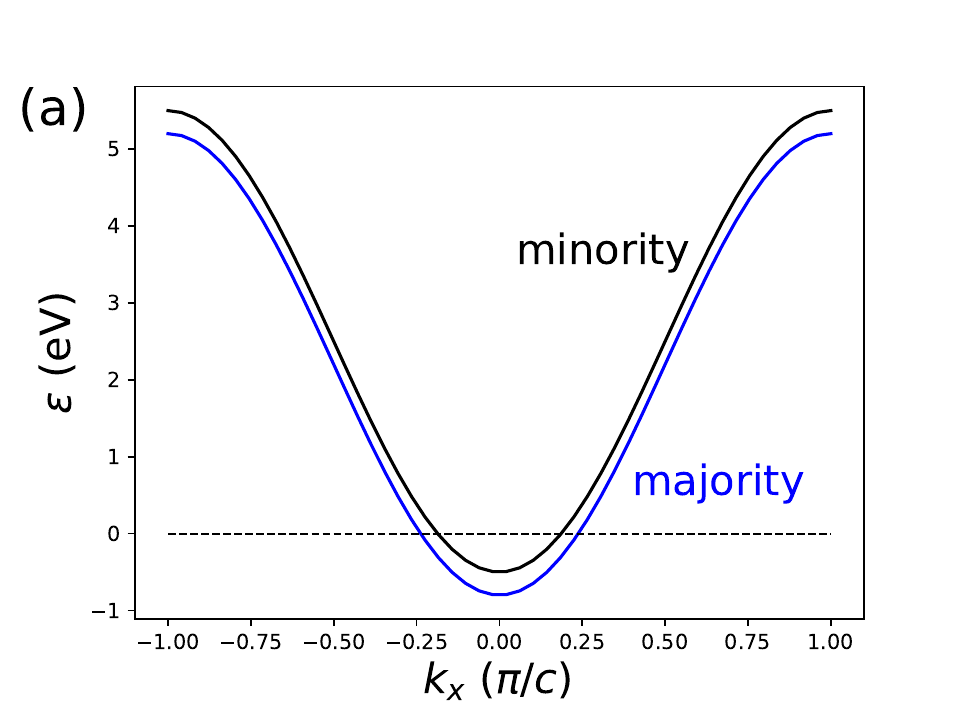}
\includegraphics[width=6cm]{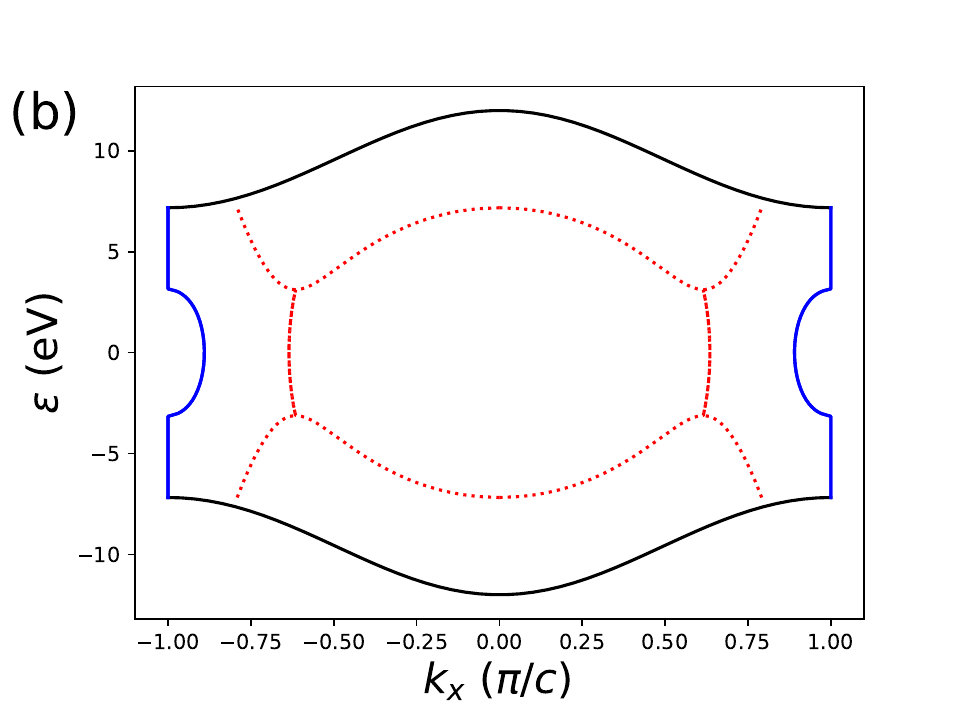}
\includegraphics[width=6cm]{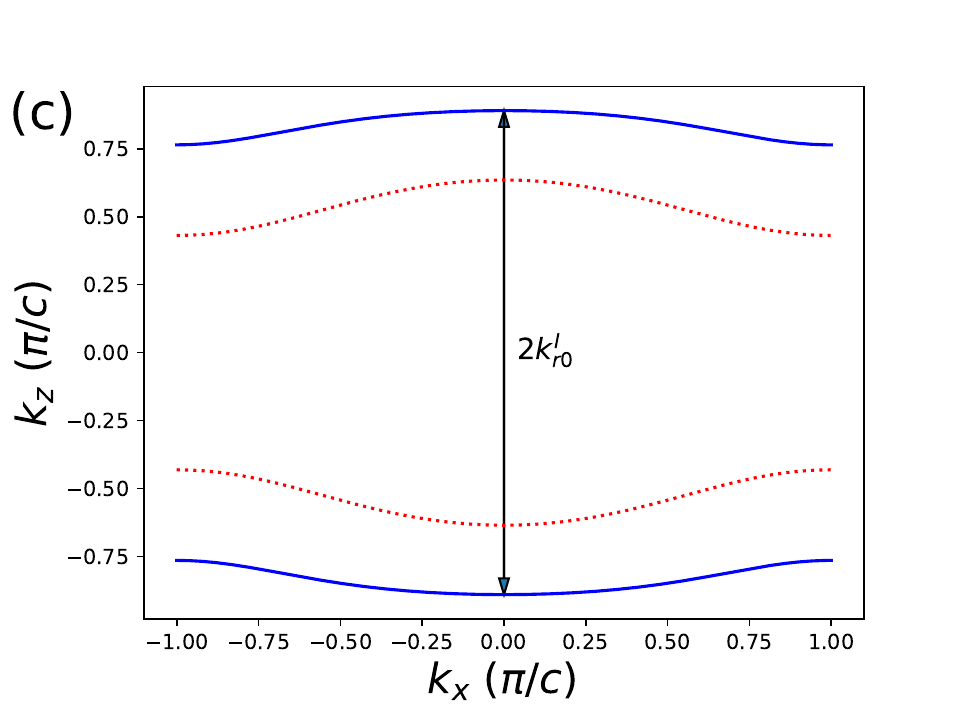}
\includegraphics[width=6cm]{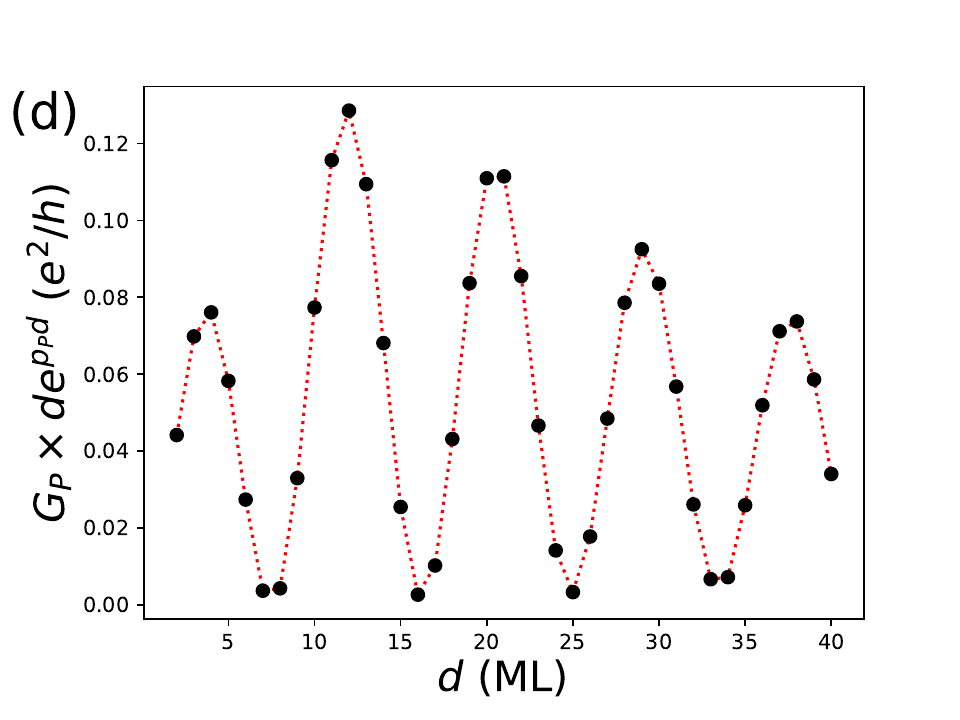}
\includegraphics[width=6cm]{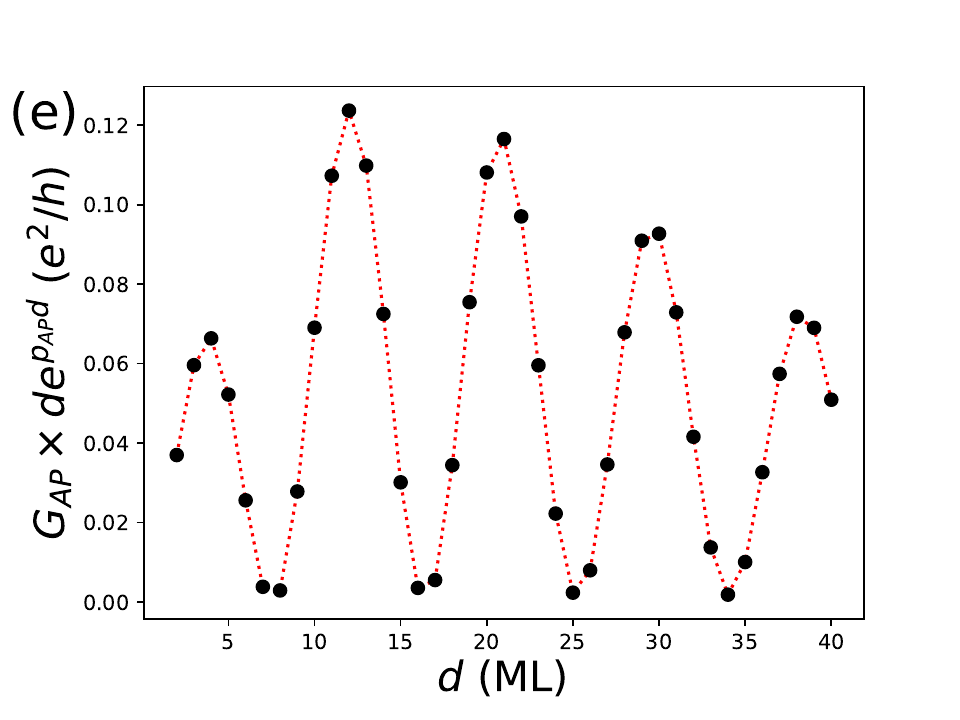}
\includegraphics[width=6cm]{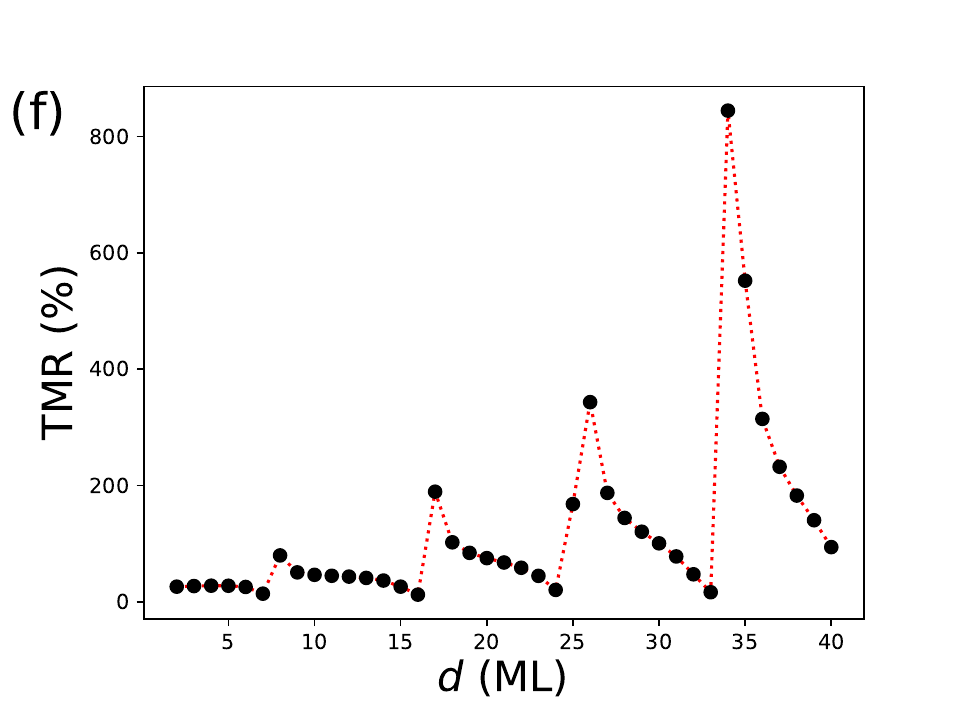}
\caption{ (a) Energy dispersion relation of the FM material
with $k_y = k_z =0$. The TB
parameters are $E_{M\uparrow}=8.2$ eV, $t_{M\uparrow}=-1.5$ eV,
$E_{M\downarrow}=8.5$ eV, and $t_{M\downarrow}=-1.5$ eV.
(b) Complex band structure of the insulator with $k_y = k_z =0$.
The solid (dotted) line is the real (imaginary) part of $k_x$. 
The TB parameters are $E_a=-5.0$ eV, $t_a=-1.0$ eV,
$E_b=5.0$ eV, $t_b=1.0$ eV, and $t_{ab}=0.8$ eV.
(c) Cross section of the complex Fermi surface along the
$k_x$ axis with $k_y=0$. The solid (dotted) line is the real 
(imaginary) part of the complex Fermi surface.
(d) Tunneling conductance of the MTJ for the parallel
magnetization. 
$p = 1.23 \pi/c$ is a fitting parameter obtained from 
extrapolation. The solid circles are the exact full-band calculation
and the dotted line is a guide for eyes.
(e) Tunneling conductance of the MTJ for the antiparallel
magnetization. The solid circles are the exact full-band calculation
and the dotted line is a guide for eyes.
$p = 1.24 \pi/c$ is a fitting parameter obtained from 
extrapolation. 
(f) TMR as a function of the barrier thickness $d$. 
The solid circles are results of the full-band calculation, and
the dotted line is a guide for eyes.
 } \label{TMR3}
\end{figure}

In Fig.~\ref{TMR3}(a), the band structure of the FM material is
similar to the previous cases, but the band structure of the insulator 
in Fig.~\ref{TMR3}(b) is totally different.
At ${\bf k}=0$, the conduction band has the maximum energy 
instead of the minimum.
Along the $k_x$ axis with $k_y = k_z = 0$, conduction energy is the lowest at $k_x = \pm \pi/c$.
The minimum energy of the conduction band is approximately 
at ${\bf k}=(0.67 \pi/c, 0.67 \pi/c, 0.67 \pi/c)$. 
The cross section of the complex Fermi surface is
shown in Fig.~\ref{TMR3}(c).
$k^{I}_{i}$, the imaginary part of $k_z$ for
the complex Fermi surface, is the largest at ${\bf k}_{\|}=0$.
$k^{I}_{i}$ decreases away from ${\bf k}_{\|}=0$ and
has a minimum value, $k^{I}_{i}$, at the extremal point
${\bf k}_{\|}={\bf k}_{\|0}=(0.67 \pi/c, 0.67 \pi/c)$.
In Fig.~\ref{TMR3}(a), 
the Fermi wavevector of the FM layer along [100] orientation is
$k_{F,x}=0.24 \pi/c$ and $0.19 \pi/c$ 
for the majority and minority spin, respectively. 
The $k^{I}_{i}$ minimum point of the insulator (${\bf k}_{\|}={\bf k}_{\|0}$)
is outside the Fermi surface of the FM layer, and
the analytical formula cannot be applied in this case.
$2k^{I}_{r0 }$ is the extremal spanning wavevector of the real part of
the complex Fermi surface in Fig.~\ref{TMR3}(c), but it does not have
much meaning because $k^{I}_{i}$ is the largest at this point and
tunneling probability is the lowest.
The tunneling conductance is shown as a function of $d$
in Fig.~\ref{TMR3}(d) for P magnetization and
in Fig.~\ref{TMR3}(e) for AP magnetization.
The solid circles are exact numerical results, and the dotted 
curve is just a guide for eyes.
As mentioned already, the analytical formula is not applied 
because the minimum point for $k^I_i$ is outside the Fermi surface 
of the FM layer.
It is still assumed that the conductance decays as $e^{-pd}/d$,
and $p$ is obtained from extrapolation.
As expected,
extrapolated values $p_P = 1.23 \pi/c$ and $p_{AP} = 1.24 \pi/c$
are smaller than $2k^I_{i}({\bf k}_{\|}=0)=1.27 \pi/c$ which is
the maximum value. 
$p_P$ is slightly smaller than $p_{AP}$, and
$G_P$  decays slower than $G_{AP}$, which is explained as follows.
 As shown
in Fig.~\ref{TMR3}(c), $k^I_i$ has the maximum value at ${\bf k}_{\|}=0$ and
decreases as $k_{\|}$ increases.
The Fermi surface of the FM layer is bigger for the majority spin
as implied in Fig.~\ref{TMR3}(a), and more electrons
with smaller $k^I_i$
contribute to tunneling conductance for P magnetization,
resulting in slower decay.
In Fig.~\ref{TMR3}(c), $2k^I_{r0}=1.78 \pi/c$ is larger 
than $\pi/c$ and the extremal spanning vector of the
real part of the complex Fermi surface is $q_{r0}=2\pi/c -2k^I_{r0}=0.22 \pi/c$.
The tunneling probability is the lowest at ${\bf k}_{\|}=0$,
and the oscillation period deviates from $2\pi/q_{r0}$.
As $k_{\|}$ increases, $k^I_r$ decreases and the spanning vector
$q_{r}=2\pi/c -2k^I_{r}$ increases. 
As the barrier thickness increases, the tunneling probability of
electrons with larger $k^I_i$ and smaller $q_r$ decays much faster, 
and the oscillation period slightly decreases.
In the same way, the oscillation period of $G_P$ is slightly shorter 
than that of $G_{AP}$.
In the FM layer, the Fermi surface is bigger for the majority spin.
Majority-spin electrons outside the minority-spin Fermi surface 
contribute only to $G_P$. They have smaller $k^I_i$ and larger $q_r$
and make the oscillation period of $G_P$ shorter.
The TMR is displayed as a function of the barrier thickness 
in Fig.~\ref{TMR3}(e).  
The solid circles are exact numerical results, and the dotted 
curve is a guide for eyes.
In addition to the oscillation,
the TMR grows rapidly as the barrier thickness increases.
It is because $G_P$ decays slower than $G_{AP}$ as explained already.
Although the difference of decay rate between $G_P$ and $G_{AP}$
is very small, the TMR increases exponentially with increasing 
the barrier thickness.

There were several experimental results for oscillatory dependence of
TMR on barrier thickness \cite{yuasa,mat,maru,Sch21,Sch22,Sch23}.
In the experiments, TMR oscillates without decaying or growing
as a function of the barrier thickness. 
The oscillation period is incommensurate with the lattice constant 
of the barrier. The oscillation period seems to be constant
with increasing the barrier thickness.
Both $G_P$ and $G_{AP}$ decay exponentially and oscillate 
with the same period. The exponential decay rate must be the
same for $G_P$ and $G_{AP}$ because TMR oscillates around  
a constant value. 
Slight difference in decay rates between $G_P$ and $G_{AP}$
would lead to exponential change of TMR as a function of 
barrier thickness.  Overall,
behavior of experimental $G_P$, $G_{AP}$, and TMR is described well by
the analytical formula derived in this paper.
The same decay rate and oscillation period of $G_P$ and $G_{AP}$ imply that
both $G_P$ and $G_{AP}$ are dominated by 
tunneling through the same electronic states in the barrier.
In the barrier, the $z$-component of the wavevector is a complex number 
which is expressed as $k^I_r ({\bf k}_{\|}) + i k^I_i ({\bf k}_{\|})$.
In the ${\bf k}_{\|}$-space of the barrier,
the most significant tunneling occurs at the extremal point 
${\bf k}_{\|}={\bf k}_{\| 0}$ where $k^I_i$ has a minimum value $k^I_{i0}$.
Given that the $z$-component of the wavevector is $k^I_{r0}+i k^I_{i0}$ 
at the extremal point of the barrier, both $G_P$ and $G_{AP}$ decay as $e^{-2k^I_{i0}d}/d$
and oscillates with period $\pi/k^I_{r0}$ 
as a function of barrier thickness $d$ \cite{period}.
In the FM layer, the electronic state is spin-dependent.
For $G_P$ and $G_{AP}$ to exhibit the same decay rate and
oscillation period, electronic states in the FM layer for
both majority and minority spin must closely align
with the electronic state with $k^I_{r0}+i k^I_{i0}$ in the barrier.
In the experimental analysis, the tunneling resistance was fitted 
using function $e^{2k^I_{i0}d}$, but improved fitting is 
anticipated with function $de^{2k^I_{i0}d}$.
In the experiments, TMR oscillation was not always sinusoidal,
with sawtooth-like oscillations being observed \cite{Sch22}.
The analytical model also suggests that the shape of the TMR oscillation 
may deviate from a sinusoidal form.


A few full-band calculations have been conducted for the tunneling 
conductance of MTJs with an MgO barrier \cite{butler,mathon2,heil}.
In the MgO barrier, ${\bf k}_{\|}=0$ is the sweet spot
where $k^I_i$ has the minimum value $k^I_{i0}$ and tunneling probability is
the highest. For the majority spin, the state with ${\bf k}_{\|}=0$
in the FM layer matches well with the electron state with 
$k^I_{i0}$ in the MgO and makes dominant contribution to $G_P$.
However, for the minority spin, the state with ${\bf k}_{\|}=0$
in the FM layer does not match with the electron state with 
$k^I_{i0}$ in the MgO due to the band symmetry.
$\mathcal{T}^L_{1 \downarrow}$,
$\mathcal{T}^R_{1 \downarrow}$, $\mathcal{T}^L_{2 \downarrow}$, and
$\mathcal{T}^R_{2 \downarrow}$ are negligible
for electrons with ${\bf k}_{\|}=0$, and
the electron state with $k^I_{i0}$ in the MgO does not
contribute to $G_{AP}$.
As a result, $G_{AP}$ decays much faster than $G_P$, and TMR increases
exponentially as a function of the barrier thickness.
Moreover, corresponding $k^I_r$ in MgO is zero at ${\bf k}_{\|}=0$
where $k^I_i$ is has the minimum value $k^I_{i0}$, and 
there are no oscillations for tunneling conductance or TMR.
Currently, the theoretical full-band calculations contradict the experimental results, 
indicating need for further investigation.

\section{Summary}

The oscillatory dependence of TMR on barrier thickness
in MTJ is investigated theoretically.
The tunneling conductance is expressed with the transmission
amplitudes at separated interfaces FM(L)/I and I/FM(R).
An analytical formula  has been developed to elucidate the relationship 
between tunneling conductance and barrier thickness.
When an electron wavefunction arrives at the left FM layer, 
it bifurcates into two tunneling waves, each possessing different complex wavevectors 
within the tunnel barrier. 
These bifurcated waves subsequently interfere with each other 
upon reconvergence at the right FM layer, 
resembling the classical optical two-slit interference phenomenon. 
The presence of complex multiple bands is a prerequisite for the oscillation 
of tunneling conductance and, subsequently, TMR oscillation.
The $z$-component of the wavevector,  $k^I_z$, in the insulator 
is expressed as $k^I_z =k^I_r + i k^I_i$ for given ${\bf k}_{\|}$
and Fermi energy.
The analytical formula is expressed with several parameters 
calculated at the extremal point, ${\bf k}_{\|}={\bf k}_{\|0}$, of the complex
Fermi surface where $k^I_{i}$ has the minimum value 
$k^I_{i0}$, and $k^I_{r}$ has the extremum value
$k^I_{r0}$.
The tunneling conductance decays as $e^{-2k^I_{i0} d}/d$
as barrier thickness $d$ increases.
The tunneling conductance oscillates at the same time, 
and the oscillation period is $\pi/k^I_{r0}$.

Numerical calculation with TB band structures
has been carried out to investigate
the validity of the analytical formula. 
A numerical scheme was presented to calculate the 
transmission coefficient of the MTJ and tunneling 
conductance.
For comparison with the analytical formula,
a spin-dependent single-band
was adopted for the FM material, and a two-band structure 
was taken for the insulator material.
The Fermi surfaces of the FM material were centered 
at ${\bf k}=0$, and it is larger for majority spin.
The $z$-component of the wavevector, $k^I_z$, in the insulator 
is complex at the Fermi level, and
the cross section of the complex Fermi surface was displayed.
Two cases were considered for the insulator. 
In the first case, the extremal point of the insulator
was located at ${\bf k}_{\|}=0$ where $k^I_i$ is the smallest.
The extremal point  ${\bf k}_{\|}=0$ is included 
in the Fermi surface of the FM material,  and
the electron states in the vicinity of ${\bf k}_{\|}=0$ 
contribute most to the tunneling conductance.
The analytical formula was in good agreement with the exact numerical calculation. 
Parameters for the analytical formula were estimated
at the extremal point ${\bf k}_{\|}=0$ and Fermi level.
$k^I_{i0}$ ($k^I_{r0}$) is 
the imaginary (real) part of $k^I_z$ in the barrier
evaluated at the extremal point ${\bf k}_{\|}=0$ and Fermi level. 
$2k^I_{r0}$ is also an extremal
spanning wavevector of the real part of 
the complex Fermi surface for the insulator
along the conduction direction.
Both $G_P$ and $G_{AP}$ decay as $e^{-2k^I_{i0} d}/d$
with increasing barrier thickness $d$, 
and the oscillation period is $\pi/k^I_{r0}$.
Dependence of the TMR on the barrier thickness
was numerically calculated, and the analytical
model agreed well with the exact calculation.
The TMR neither grows nor decays, and
oscillates with the same period $\pi/k^I_{r0}$ as conductance.
The oscillation is not sinusoidal and
the oscillation shape varies a lot each time.
The peak-to-valley difference of the TMR was about 80 \%.
It was shown that the complex wavevector of the barrier 
may give a rise to the significant TMR oscillation as a function 
of the barrier thickness.
The most important two parameters are determined from
the complex Fermi surface of the insulator material.
The oscillation period is related to 
the extremal spanning wavevector of the
real part of the insulator Fermi surface along
the conduction direction, 
while the corresponding imaginary part of $k_z$
has the smallest value $k^I_{i0}$.

In the second case, $k^I_i$ of the barrier
has the maximum value at ${\bf k}_{\|}=0$, and
the extremal point, where $k^I_i$  has the smallest value $k^I_{i0}$, is
far away from ${\bf k}_{\|}=0$.
The extremal point is outside the FM Fermi surface,
and electron state with $k^I_{i0}$ is not involved in the tunneling conductance. 
Decay rate of $G_P$ is smaller than that of $G_{AP}$ due to the size difference 
of the FM Fermi surfaces between majority and minority spin.
Although the analytical formula is not directly 
applicable in this case, the decay of the tunneling conductance
is fitted to some extend by $e^{-q d}/d$ with adjustable fitting parameter $q$.
The oscillation period of the tunneling conductance changes slightly as the barrier thickness increases.
The TMR grows exponentially because $G_{AP}$ decays faster than $G_P$. 
The TMR oscillates at the same time, and the oscillation amplitude also grows exponentially. 

The analytical formula derived in this paper is consistent with
TMR behavior observed in the experiments.
In the experiments, the TMR oscillates around
a constant value as a function of the barrier thickness. 
Both $G_P$ and $G_{AP}$ oscillate with the
same period in addition to the same exponential decay rate.
It means that, for both majority and minority spin,
the tunneling is dominated
by the electrons with $k_{\|}$ in the vicinity of the same 
extremal point of the barrier where $k^I_i$ has the smallest value
and $k^I_r$ has an extremum value.
However, other full-band calculations suggest differently. 
In the MTJ with the MgO barrier, the extremal point of the barrier is
located at $k_{\|}=0$ where $k^I_i$ is has the smallest value. 
According to other full-band calculations, 
the electrons with the minority spin in the FM layer do not match 
with those in the barrier at $k_{\|}=0$.
As a result, $G_{AP}$ decays faster than $G_P$ and TMR 
grows as functions of the barrier thickness. 
Moreover, for the barrier, $k^I_r$ is zero at  $k_{\|}=0$, 
and oscillation of the tunnel conductance has not been explained.
This discrepancy requires further investigation.

\begin{acknowledgments}
This work was supported by INHA UNIVERSITY Research Grant.
\end{acknowledgments}

\newpage

\newpage

\newpage

\newpage

\newpage

\newpage

\begin{thebibliography}{99}
\bibitem{period} If $k^I_{r0}$ is larger than $\pi /2c$,
the oscillation period is $2\pi /(2\pi /c - 2k^I_{r0}) $.
\end{thebibliography}

%

\end{document}